\documentclass[12pt]{article}
\usepackage[margin=1in]{geometry}

\usepackage{amsmath,amsthm,amssymb}

\usepackage{float}
\usepackage{algpseudocode}
\usepackage{enumitem}
\usepackage[utf8]{inputenc} 
\usepackage[T1]{fontenc}    
\usepackage{hyperref}       
\usepackage{url}            
\usepackage{booktabs}       
\usepackage{amsfonts}       
\usepackage{nicefrac}       
\usepackage{microtype}      
\usepackage{xcolor}         
\usepackage{subcaption}
\usepackage{graphicx}
\usepackage{caption}
\usepackage{algorithm}
\usepackage{wrapfig}
\usepackage[noblocks]{authblk}

\title{Hierarchical Bias-Driven Stratification for Interpretable Causal Effect Estimation}

\begin{document}

\author[1, 2]{Lucile Ter-Minassian}
\author[1]{Liran Szlak} %
\author[1]{Ehud Karavani} 
\author[2]{Chris Holmes} 
\author[1]{Yishai Shimoni\footnote{Correspondence: yishais@il.ibm.com}}

\affil[1]{\footnotesize IBM Research, Israel}
\affil[2]{\footnotesize Department of Statistics, University of Oxford}

\date{\today}

\maketitle

\begin{abstract}
Interpretability and transparency are essential for incorporating causal effect models from observational data into policy decision-making. They can provide trust for the model in the absence of ground truth labels to evaluate the accuracy of such models. 
To date, attempts at transparent causal effect estimation consist of applying post hoc explanation methods to black-box models, which are not interpretable. 
Here, we present BICauseTree: an interpretable balancing method that identifies clusters where natural experiments occur locally. Our approach builds on decision trees with a customized objective function to improve balancing and reduce treatment allocation bias. Consequently, it can additionally detect subgroups presenting positivity violations, exclude them, and provide a covariate-based definition of the target population we can infer from and generalize to. We evaluate the method's performance using synthetic and realistic datasets, explore its bias-interpretability tradeoff, and show that it is comparable with existing approaches. 
\end{abstract}

\section{Introduction}

The goal of causal inference is to estimate the strength of a pre-specified intervention on some outcome of interest.
It is therefore a valuable tool for data-driven decision-making in public health, economics, political science, and more.
These are high-stakes, high-visibility domains in which data-based conclusions should ideally be understood by both the policy makers and the general public, which lack the expertise in the inner workings of statistical models.
This, in turn, incentivizes the use of \emph{interpretable} models for estimating causal effects.

Following Rudin (2019) \cite{rudin2019stop}, \emph{interpretability} means that each decision in the algorithm is inherently explicit and traceable, contrasting with \emph{explainability} where decisions are justified post-hoc using an external model.
Therefore, when involving laypeople, interpretable models may often be preferred over explainable ones, 
as the latter might require understanding the assumptions and implications of an additional explainability model on top of the effect estimation model. 
Simplicity may incur better trustworthiness. 

Interpretable or not, modeling causal effects requires estimating the \emph{potential outcomes} - the outcome a person would experience under every possible intervention \cite{rubin1974estimating}.
However, at any given time, a person can either be treated or not, but not both.
This ``fundamental problem of causal inference'' \cite{holland1986statistics} implies we never have ground truth labels and thus can never evaluate the accuracy of estimated causal effects.
Furthermore, in nonexperimental settings where treatment is not assigned randomly, groups that do or do not receive treatment may not be comparable in their attributes, and such attributes can influence the outcome too, introducing confounding bias.
These two difficulties can harm the trustworthiness of data-driven conclusions perceived by both policy makers and the general public.
However, by being explicitly transparent, interpretable models allow a new way to evaluate causal models by judging the sensibility of the conclusions they made and whether they adhere to one's common sense.

In this paper, we introduce BICauseTree: Bias-balancing Interpretable Causal Tree, an interpretable balancing method for observational data with binary treatment that can handle high-dimensional datasets. We use a binary decision tree to stratify on imbalanced covariates and identify subpopulations with similar propensities to be treated, when they exist in the data. The resulting clusters act as local naturally randomized experiments. This newly formed partition can be used for effect estimation, as well as propensity score or outcome estimation. Our method can further identify positivity-violating regions of the data space, i.e.,\ subsets of the population where treatment allocation is highly unbalanced. By doing so, we generate a transparent, covariate-based definition of the target (``inferentiable'') population, i.e.\ the population with sufficient overlap for inferring a causal effect.

Our contributions are as follows:
\begin{enumerate}[noitemsep, leftmargin=0.8cm]
    \item Our BICauseTree method can identify ``natural experiments'' i.e.\ subgroups with lower treatment imbalance, when they exist. 
    \item BICauseTree compares with existing methods for causal effect estimation in terms of bias while maintaining interpretability. Estimation error and consistency of the clusters show good robustness to subsampling in both our synthetic and realistic benchmark datasets.
    \item Our method provides users with a built-in prediction abstention mechanism for covariate spaces lacking common support. We show the value of defining the inferentiable population using a clinical example with matched twins data. 
    \item The resulting tree can further be used for propensity or outcome estimation.
    \item We release open-source code with detailed documentation for implementation of our method, and reproducibility of our results.
\end{enumerate}

\section{Related work}\label{sec:related_work}
\subsection{Effect estimation methods}
Causal inference provides a wide range of methods for effect estimation from data with unbalanced treatment allocation. There are two modelling strategies: modelling the treatment using the covariates to balance the groups, and modelling the outcome directly using the covariates and treatment assignment. 

In balancing methods, such as matching or weighting methods, the data is pre-processed to create subgroups with lower treatment imbalance or  ``natural experiments''. 
\emph{Matching} methods consist of clustering similar units, based on some distance metric (Euclidean/Mahalanobis distance, nearest neighbor search, etc), from the treatment and control groups to reduce imbalance. However, as the notion of distance becomes problematic in high dimensional spaces, covariate-based matching tends to become ineffective \cite{beyer1999nearest}. 
\emph{Weighting} methods aim at balancing the covariate distribution across treatment groups, with Inverse Probability Weighting (IPW) \cite{horvitz1952generalization} being the most popular approach. Sample weights are the inverse of the estimated \emph{propensity} scores, i.e.\ the probability of a unit to be assigned to its observed group. However, extreme IPW weights can increase the estimation variance.

Contrastingly, in \emph{adjustment} methods the causal effect is estimated from regression outcome models where both treatment and covariates act as predictors of the outcome. These regressions can be fitted through various methods like linear regression \cite{imbens2015causal}, neural networks \cite{shi2019adapting, shalit2017estimating}, or tree-based models \cite{athey2016recursive, kunzel2019metalearners}.

BICauseTree is a \emph{balancing} method, i.e., a data-driven mechanism for achieving conditional exchangeability. 
Nonetheless, it can be combined with other methods to achieve superior results.
Either as propensity models in established doubly robust methods \cite{kang2007demystifying},
or by incorporating arbitrary causal models at leaf nodes (similar to regression trees with linear models at leaf nodes \cite{quinlan1992learning}).

\subsection{Positivity violations}
Causal inference is only possible under the \emph{positivity} assumption, which requires covariate distributions to overlap between treatment arms. Thus, positivity violations (also referred to as no overlap) occur when certain subgroups in a sample do not receive one of the treatments of interest or receive it too rarely \cite{karavani2019discriminative}. Overlap is essential as it guarantees data-driven outcome extrapolation across treatment groups. Having no common support means there are subjects in one group with no counterparts from the other group, and, therefore, no reliable way to pool information on their outcome had they been in the other group. Non-violating samples are thus the only ones for which we can guarantee some validity of the inferred causal effect. 

There are three common ways to characterize positivity. The most common one consists in estimating propensity scores and excluding the samples associated with extreme values (known as ``trimming'') \cite{petersen2012diagnosing}. The threshold for propensity scores can be set arbitrarily or dynamically \cite{crump2009dealing}. However, since samples are excluded on the basis of their propensity scores and not their covariate values, these methods lack interpretability about the excluded subjects and how it may affect the target population on which we can generalize the inference. Thus, other methods have been developed to overcome this challenge by characterizing the propensity-based exclusion \cite{oberst2020characterization, wolf2021positivity, ackerman2020detection}. Lastly, the third way tries to characterize the overlap from covariates and treatment assignment directly, without going through the intermediate propensity score e.g.\ PositiviTree \cite{karavani2019discriminative}. In PositiviTree, a decision tree classifier is fitted to predict treatment allocation. In contrast to their approach, BICause Tree implements a  tailor-made optimization function where splits are chosen to maximize balancing in the resulting sub-population, whereas PositiviTree uses off-the-shelf decision trees maximizing separation. 
Ultimately, the above-mentioned methods for positivity identification and characterization are model agnostic. In our model, BICauseTree, positivity identification \emph{and characterization} are inherently integrated into the model, and effect estimation comes with a built-in interpretable abstention prediction mechanism.

\subsection{Interpretability and causal inference}
A predominant issue in existing effect estimation methods is their lack of interpretability. A model is considered as \emph{interpretable} if its decisions are inherently transparent \cite{rudin2019stop}. Examples of interpretable models include decision trees where the decision can be recovered as a simple logical conjunction. Contrastingly, a model is said to be \emph{explainable} when its predictions can be justified a-posteriori by examining the black-box using an additional ``explanation model''. Popular post-hoc explanation models include Shapley values \cite{lundberg2017unified} or LIME \cite{ribeiro2016model}. However, previous works have shown that existing explainability techniques lack robustness and stability \cite{mittelstadt2019explaining}. Further, the explanations provided by explanation models inevitably depend on the black-box model's specification and fitness. 
Given that explanation models only provide unreliable justifications for black-box model decisions, a growing number of practitioners have been advocating for intrinsically interpretable predictive models \cite{rudin2019stop}. We further claim that causal inference, and in particular effect estimation,  should be \emph{interpretable} as it assists high-stake decisions affecting laypeople. 

Causal Trees (CT) \cite{athey2016recursive} are another tree-based model for causal inference that (i) leverages the inherent interpretability of decision trees, and (ii) has a custom objective function for recursively splitting the data. Although both utilize decision trees, BICauseTree and CT serve distinct purposes. BICauseTree splits are optimized for balancing treatment \emph{allocation} while CT splits are optimized for balancing treatment \emph{effect}, under assumed exchangeability. In other words, CT \emph{assumes} exchangeability while BICauseTree \emph{finds} exchangeability. 
As such, our approach is only suited for ATE estimation while CT is better suited for Conditional Average Treatment Effect estimation \cite{athey2016recursive}. Furthermore, in practice, causal effects are often averaged over multiple trees into a Causal \emph{Forest} \cite{wager2018estimation,athey2019generalized} that is no longer interpretable, and users are encouraged to use post-hoc explanation methods \cite{econml}.

In addition to effect estimation, positivity violations characterization should also be interpretable for downstream users, such as policy makers. 
Discarding samples can hurt the external validity of any result, as there can be structural biases leading to a subpopulation being excluded. 
Therefore, interpretable characterization of the overlap in a study can help policy makers better assess to whom they expect the study results to apply \cite{oberst2020characterization,karavani2019discriminative}.
BICauseTree generates a covariate-based definition of the violating subpopulation. In other words, we can claim which target population our estimate of the Average Treatment Effect applies to.

\section{BICauseTree}\label{sec2}

\subsection{Formal causal model}\label{sec:causal_inf}
We consider a dataset of size $n$ where we note each individual sample $\left(X_i, T_i, Y_i\right)$ with $X_i \in \mathbb{R}^d$ is a covariate vector for sample $i$ measured prior to binary treatment allocation $T_i$. In the potential outcomes framework \cite{rubin1973use}, $Y_i(1)$ is the outcome under $T_i=1$, and $Y_i(0)$ is the analogous outcome under $T_i=0$. Then, assuming the consistency assumption, the observed outcome is defined as $Y_i=T_i Y_i(1)+\left(1-T_i\right) Y_i(0)$. In this paper, we focus on estimating the average treatment effect (ATE), defined as:  $\mathrm{ATE}=\mathbb{E}[Y(1)-Y(0)]$.

\subsection{Motivation}\label{paragraph:motivations}
We introduce a method for balancing observational datasets with the goal of estimating the causal effect in a subpopulation with sufficient overlap. Our goals are: (i) unbiased estimation of causal effect, (ii) interpretability of both the balancing and positivity violation identification procedures, (iii) ability to handle high-dimensional datasets. Our approach utilizes the Absolute Standardized Mean Difference (ASMD) \cite{austin2009balance} frequently used for assessing potential confounding bias in observational data. Note that our balancing procedure is entirely interpretable, although it can be used in combination with arbitrary black-box outcome models or propensity score models. Finally, our method generates a covariate-based definition of the target population on which we make inference. As such, it is tailored to sensitive domains where inference should be restricted to subpopulations with reasonable overlap. We envision this method to be used in fields such as epidemiology, econometrics, medicine, and political science. 

\subsection{Algorithm}
The intuition for our algorithm is that, by partitioning the population to maximize treatment allocation heterogeneity, we may be able to find subpopulations that are natural experiments. We recursively partition the data according to the most imbalanced covariate between treatment groups. Using decision trees makes our approach transparent and non-parametric.

\paragraph{Splitting criterion} 
The first step of our algorithm is to split the data until some stopping criterion is met. The tree recursively splits on the covariate that maximize treatment allocation heterogeneity. To do so, we compute the Absolute Standardized Mean Difference (ASMD) for all covariates and select the covariate with the highest absolute value. The ASMD for a variable $X_j$ is defined as: \\\\
\centerline{$\mathrm{ASMD_j}=\frac{\vert\mathbb{E}[X_j|T=1]-\mathbb{E}[X_j|T=0]\vert}{\sqrt{Var([X_j|T=1]) + Var([X_j|T=0])}}$} \\\\
Since we assume all confounders are observed, The reason for choosing the one with the highest ASMD is that it is most likely to cause the most confounding bias, and therefore adjusting for it will likely minimize the residual confounding bias (Figures \ref{fig:random_split} and \ref{fig:asmd_adjustment}). 
Once that next splitting covariate $j_{max}$ is chosen, we want to find a split that is most associated with treatment assignment, so that we may control for the effect of confounding. The tree finds the optimal splitting value by iterating over covariate values $x_{j_{max}}$ and taking the value associated with the lowest \textit{p}-value according to a Fisher's exact test or a $\chi^2$ test, depending on the sample size.

\paragraph{Stopping criterion} 
Tree building stops when either: (i) the maximum ASMD is below some threshold, (ii) the minimum treatment group size falls below some threshold  (iii) the total population fall below a threshold, or (iv) a maximum tree depth is reached. All thresholds are user-defined hyperparameters. 

\paragraph{Pruning procedure} 
Once the stopping criterion is met in all leaf nodes, the tree is pruned. A multiple hypothesis test correction is first applied on the \textit{p}-values of all splits. Following this, the splits with significant \textit{p}-values or with at least one split with significant \textit{p}-value amongst their descendants are kept. Ultimately, given that ASMD reduction may not be monotonic, pruning an initially deeper tree allows us to check if partitioning more renders unbiased subpopulations. The implementation of the tree allows for user-defined multiple hypothesis test correction, with current experiments using Holm-Bonferonni \cite{holm1979simple, abdi2010holm}. The choice of the pruning and stopping criterion hyperparameters will guide the bias/variance trade-off of the tree. Deeper trees may have more power to detect treatment effect while shallower trees will be more likely to have biased effect estimation.

\paragraph{Positivity violation filtering} 
Lastly, we evaluate the overlap in the resulting set of leaf nodes to identify those where inference is possible. 
We check for treatment balance based on a user-defined overlap estimation method, 
with the default method being the Crump procedure \cite{crump2009dealing}. 
The positivity-violating leaf nodes are tagged and used for inference abstention mechanism, 
i.e.\ inference will be restricted to non-violating leaves. 

\paragraph{Estimation}
Once a tree is contracted, it can be used to estimate both counterfactual outcomes and propensity scores. For each leaf, using the units propagated to that leaf, we can model the counterfactual outcome by taking the average outcome of those units in both treatment groups. Alternatively, we can fit any arbitrary causal model (e.g., IPW or an outcome regression) to obtain the average counterfactual outcomes in that leaf. The ATE is then obtained by averaging the estimation across leaves. Similarly, we can estimate the propensity score in each leaf by taking the treatment prevalence or using any other estimator (e.g., logistic regression).

\paragraph{Code and implementation}
BICauseTree code is released open-source under: \url{https://github.com/IBM-HRL-MLHLS/BICause-Trees}. Our flexible implementation allows users to define a stopping criteria as well as a multiple hypothesis correction method. BICauseTree adheres to \texttt{causallib}'s API, and can accept various outcome and propensity models.

\begin{algorithm}[H]
\begin{algorithmic}
\caption{\textbf{BICauseTree}}
\State{\textbf{Inputs:} root node $N_0$, $X$, $T$, $Y$ }
\State{Call \textit{Build subtree}($N_0$, $X$, $T$, $Y$)}
\State{Do multiple hypothesis test correction on all split \textit{p}-values}
\State{\textbf{Pruning procedure}: keep splits with either (i) a significant \textit{p}-value or (ii) at least one descendant with a significant \textit{p}-value}
\State{Mark leaf nodes that violate positivity criterion}
\end{algorithmic}
\end{algorithm}
\vspace{-0.2cm}
\begin{algorithm}[]
\begin{algorithmic}
\caption{\textbf{Build subtree}}
\State{\textbf{Inputs:} current node $N$, $X$, $T$, $Y$ }
\If{Stopping criteria not met}
    \State {Find and record in $N$ the covariate with maximum ASMD: $max ASMD := max_{i}(ASMD_i)$} 
    \State{Find and record in $N$ the split value with the lowest \textit{p}-value according to a Fisher test/$\chi^2$ test}
    \State{Record the \textit{p}-value for this split in $N$}
    \State {Split the data ${X, T, Y}$ into ${X_{left}, T_{left}, Y_{left}}$ and ${X_{right}, T_{right}, Y_{right}}$ according to $N$'s splitting covariate and value} 
\State{Add two child nodes to $N$: $N_{left}$ and $N_{right}$}
\State {Call \textit{Build subtree}($N_{left}$, $X_{left}$, $T_{left}$, $Y_{left}$)} 
\State {Call \textit{Build subtree}($N_{right}$, $X_{right}$, $T_{right}$, $Y_{right}$)}
\EndIf
\end{algorithmic}
\end{algorithm}

\section{Experiment and results}\label{sec:exp}

\subsection{Experimental settings}\label{paragraph:baseline_comp}
\vspace{-0.2cm}
In all experiments--unless stated otherwise--the data was split into a training and testing set with a 50/50 ratio. The training set was used for the construction of the tree and for fitting the outcome models in leaf nodes, if relevant. Causal effects are estimated by taking a weighted average of the local treatment effects in each subpopulation. At the testing phase, the data is propagated through the tree, and potential outcomes are evaluated using the previously fitted leaf outcome model. We performed 50 random train-test splits, which we will refer to as \emph{subsamples} to avoid confusion with the tree partitions. For each subsample, effects are only computed on the non-violating samples of the population. In order to maintain a fair comparison, these samples are also excluded from effect estimation with other models and with ground truth. 

\paragraph{Baseline comparisons}
We compare our method to double Mahalanobis Matching, Inverse Probability Weighting (IPW), and Causal Tree (CT). In Mahalanobis Matching \cite{rubin1980bias, stuart2010matching}, the nearest neighbor search operates on the Mahalanobis distance: $d\left(X_i, X_j\right)=\left(X_i-X_j\right)^T \Sigma^{-1}\left(X_i-X_j\right)$, where $\Sigma$ is alternatively the estimated covariance matrix of the control and treatment group dataset. In Inverse Probability Weighting \cite{horvitz1952generalization}, a propensity score model estimates the individual probability of treatment conditional on the covariates. The data is then weighted by the inverse propensities $P(T=t_i \mid X=\boldsymbol{x}_i)^{-1}$ to generate a balanced pseudo-population. In Causal Tree \cite{athey2016recursive}, the splitting criterion optimizes for treatment \emph{effect} heterogeneity (see section \ref{sec:further_related_work} for further details). We use a Causal Tree and not a Causal Forest to compare to an estimator which is equally interpretable as our estimator. 
We also compare our results to an unadjusted marginal outcome estimator, which will act as our ``dummy'' baseline model. As using a single Causal Tree for our interpretability goal gives rise to high estimation bias, Causal Tree was excluded from the main manuscript for scaling purposes. We refer the reader to sections \ref{sec:further_sim}, \ref{sec:further_twins} and \ref{sec:further_acic} for a comparison with CT. For synthetic experiments, we use the simplest version of our tree which we term BICauseTree(Marginal) where the effect is estimated by taking average outcomes in leaf nodes. For real-world experiments, we compare BICauseTree(Marginal) with BICauseTree(IPW), an augmented version in which an IPW  model is fitted in each leaf node. To compare estimation methods, we compute the difference between the estimated ATE and the true ATE for each subsample (or train-test partition) and display the resulting distribution of estimation biases in a box plot. Further experimental details, including hyperparameters, can be found in the Appendix under section \ref{sec:exp_details}.

\subsection{Synthetic datasets}

We first evaluate the performance of our approach on two synthetic datasets. We first demonstrate BICauseTree's ability to identify subgroups with lower treatment imbalance on a dataset which we will refer to as the ``natural experiment dataset''. We further exemplify BICauseTree's identification of positivity-violating samples on a dataset we refer to as the ``positivity violations dataset''. Due to the interaction-based nature of the data generation procedure, we additionally compare our approach to an IPW estimator with a Gradient Boosting classifier propensity model, referred to as IPW (GBT) in both synthetic experiments. This choice ensures a fair comparison across estimators.

\paragraph{Identifying natural experiments}\label{sec:sim1}
For the natural experiment dataset, we used a Death outcome $D$, binary treatment of interest $T$ and two covariates: Sex $S$ and Age $A$. We defined four sub-populations, each constituting a natural experiment with a truncated normal propensity distribution centered around a pre-defined mean and variance (see Section \ref{sec:natural_exp}). Then, individual treatment propensities were sampled from the corresponding distribution, and observed treatment values were sampled from a Bernoulli parameterized with the individual propensities. No positivity violation was modeled. Ultimately, $X=(S, A) \in \mathbb{R}^2$ is the vector of covariate values. Sample size is $n=20,000$. The marginal distribution of covariates follows: $S \sim \operatorname{Ber}(0.5), A\sim \mathcal{N}\left(\mu , \sigma^2\right)$ where $\mu = 50, \sigma = 20$.

\begin{figure*}[t!]
    \centering
    \begin{subfigure}[t]{0.49\textwidth}
        \centering
        \includegraphics[width=\textwidth]{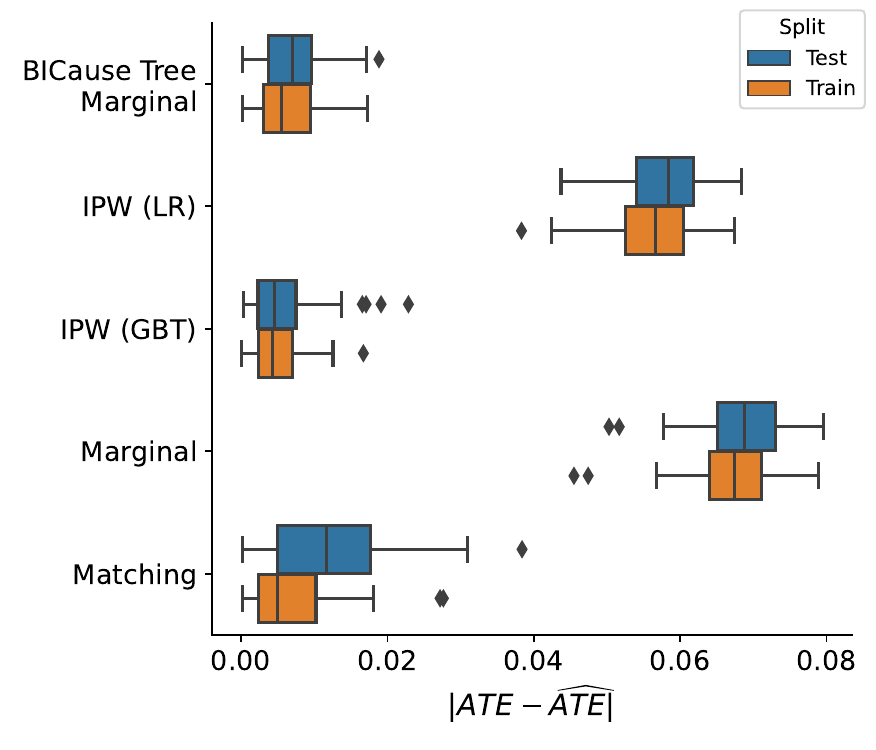}
    \end{subfigure}%
    ~ 
    \begin{subfigure}[t]{0.49\textwidth}
        \centering
        \includegraphics[width=\textwidth]{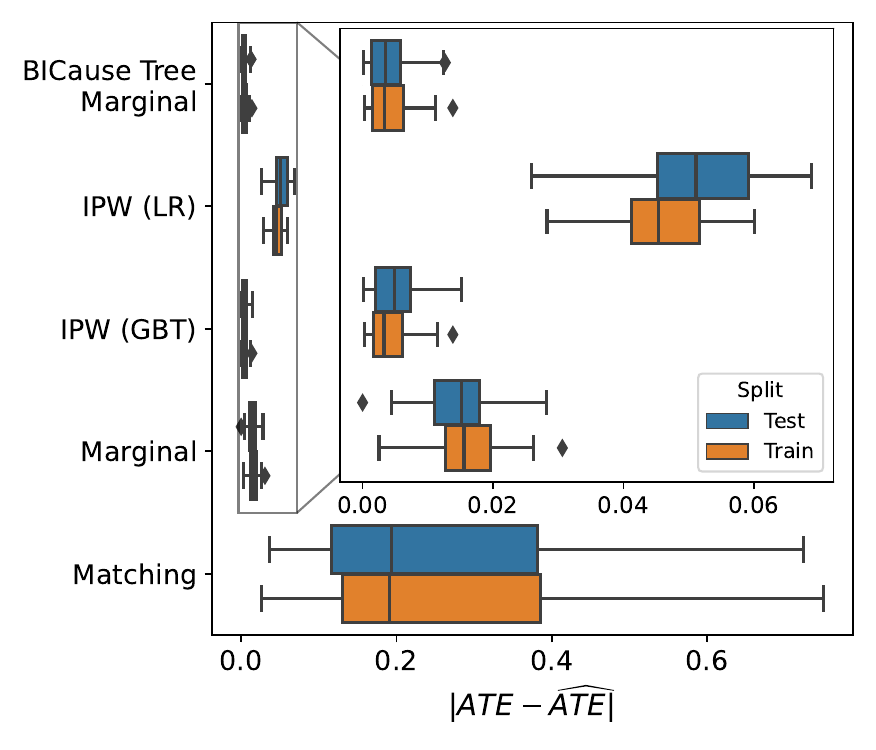}
    \end{subfigure}
    \caption{Estimation bias for (left) the natural experiment dataset and (right) the positivity violations dataset across 50 subsamples, with $N = 20,000$. In the natural experiment (left), on top of being transparent, BICauseTree has lower bias in causal effect estimation compared to all other methods, excluding IPW(GBT) which has comparable performance. In the positivity violation experiment (right), after filtering violating samples the effect estimation by BICauseTree remains unbiased and with low variance.}
    \label{fig:sim_dataset_diff}
\end{figure*}

Figure \ref{fig:tree_entire_sim1} in \ref{sec:further_sim1} shows the partition obtained from training BICauseTree on the entire dataset. Our tree successfully identifies the subpopulations in which a natural experiment was simulated. Figure \ref{fig:sim_dataset_diff} (left) shows the estimation bias across subsamples. In addition to being transparent, BICauseTree has lower bias in causal effect estimation compared to all other methods, excluding IPW(GBT) which has comparable performance. Despite its higher estimation variance, Matching has low bias, probably due to covariate space being well-posed and low-dimensional. Contrastingly, the logistic regression in IPW(LR) is not able to model treatment allocation as the true propensities are generated from a noisy piecewise constant function of the covariates resulting in a threshold effect that explains its poor performance. The non-parametric, local nature of both Matching and BICauseTree thus contrasts with the parametric estimation by IPW(LR). Further results on the BICauseTree's calibration and covariate partition can be found in the Appendix, under section \ref{sec:further_sim1}.

\paragraph{Identifying positivity violations}\label{sec:sim2}

For the positivity violations dataset, we consider a synthetic dataset with a Death outcome $D$, a binary treatment of interest $T$, and three Bernoulli covariates --Sex $S$, cancer $C$ and arrhythmia $A$-- such that $X=(S, C, A)$ (see Section \ref{sec:positivity} for further details). As for the natural experiment dataset, we modeled treatment allocation with stochasticity by sampling propensities from a truncated gaussian distribution first. Treatment allocation was simulated to ensure that overlap is very limited in two subpopulations: females with no cancer and no arrhythmia are rarely treated, while males with cancer and arrhythmia are almost always treated. 
Figure \ref{fig:partition_positivity} in \ref{sec:further_sim2} shows the partition obtained from training BICauseTree on the entire dataset, confirming that BICauseTree excludes the subgroups where positivity violations were modeled. On average, 67.1\% of the cohort remained after positivity filtering with very little variability across subsamples. Thanks to the interpretable nature of our method, we are able to identify these subgroups as a region of the covariate space. As seen in Figure~\ref{fig:sim_dataset_diff} (right), after filtering violating samples the effect estimation by BICauseTree remains unbiased and with low variance. Our estimator compares with IPW(GBT) while being interpretable. The IPW(LR) estimator is more biased than BICauseTree. This may be due to the extreme weights in the initial overall cohort. In spite of filtering samples from regions with lack of overlap--as defined by BICauseTree--the remaining propensity weights may be biased, which would ultimately induce a biased effect estimation. Estimation variance is comparable across methods, except for Matching which is both more biased and has higher variance than all other estimators. Further results on the BICauseTree's calibration and covariate partition can be found in the Appendix, under section \ref{sec:further_sim2}.

\subsection{Realistic datasets}\label{sec:realistic_datasets}

\paragraph{Causal benchmark datasets}\label{sec:twins_acic}
 We use two causal benchmark datasets to show the value of our approach. The twins dataset illustrates the high applicability of our procedure to clinical settings. It is based on real-world records of $N = 11,984$ pairs of same-sex twin births and has 75 covariates. It tests the effect of being born the heavier twin (i.e.\ the treatment) on death within one year (i.e.\ the outcome). We use the dataset generated by \textit{Neal et. al} \cite{neal2020realcause}, which simulates an observational study from the initial data by selectively hiding one of the twins with a generative approach. We also ran our analysis on the \textit{2016 Atlantic Causal Inference Conference} (ACIC) semisynthetic dataset with simulated outcomes \cite{hahn2019atlantic}. For ACIC, given that trees are data greedy, and due to the smaller sample size ($N=4,802$) relative to the number of covariates ($d = 79$), the models were trained on 70\% of the dataset.

\begin{figure*}[t!]
    \centering
    \begin{subfigure}[t]{0.49\textwidth}
        \centering
        \includegraphics[width=\textwidth]{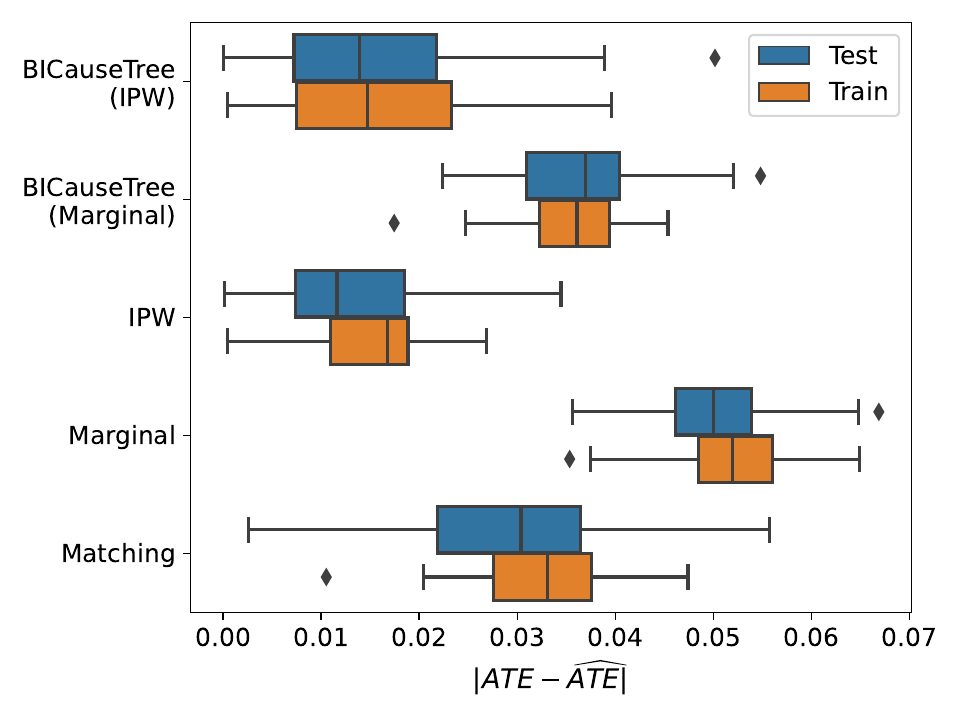}
    \end{subfigure}%
    ~ 
    \begin{subfigure}[t]{0.49\textwidth}
        \centering
        \includegraphics[width=\textwidth]{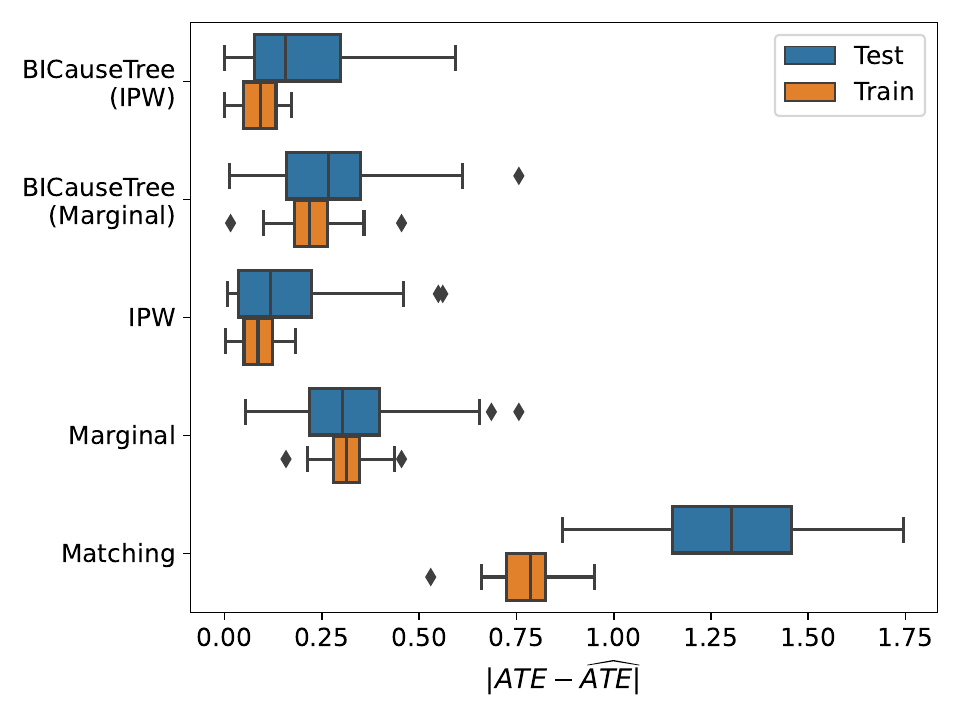}
    \end{subfigure}
    \caption{Estimation bias for (left) the twins dataset and (right) the ACIC dataset across 50 subsamples, with $N = 20,000$ excluding positivity-violating leaf nodes. For the twins dataset (left), the BICauseTree(Marginal) estimator is less biased than the marginal estimator. Augmenting our tree with an IPW outcome model (BICauseTree(IPW)) further decreases estimation bias, making it comparable with IPW wrt both bias and estimation variance. For the ACIC dataset (right), both BICauseTree models compare with IPW wrt estimation bias and variance.}
    \label{fig:real_dataset_diff}
\end{figure*}

\paragraph{Effect estimation}
 Figure \ref{fig:real_dataset_diff} (left) shows the distribution of estimation bias across subsamples on the twins dataset, compared to the baseline models. Here, our BICauseTree(Marginal) estimator is less biased than the marginal estimator. Augmenting our tree with an IPW outcome model (BICauseTree(IPW)) further decreases estimation bias, making it comparable with IPW, both w.r.t bias and estimation variance. Figure \ref{fig:real_dataset_diff} (right) compares the estimation bias on the ACIC dataset. Here, both BICauseTree models compare with IPW in estimation bias and variance. 
 
\paragraph{Bias-interpretability tradeoff}
\begin{figure}[ht]
\centering
\includegraphics[width=0.9\textwidth]{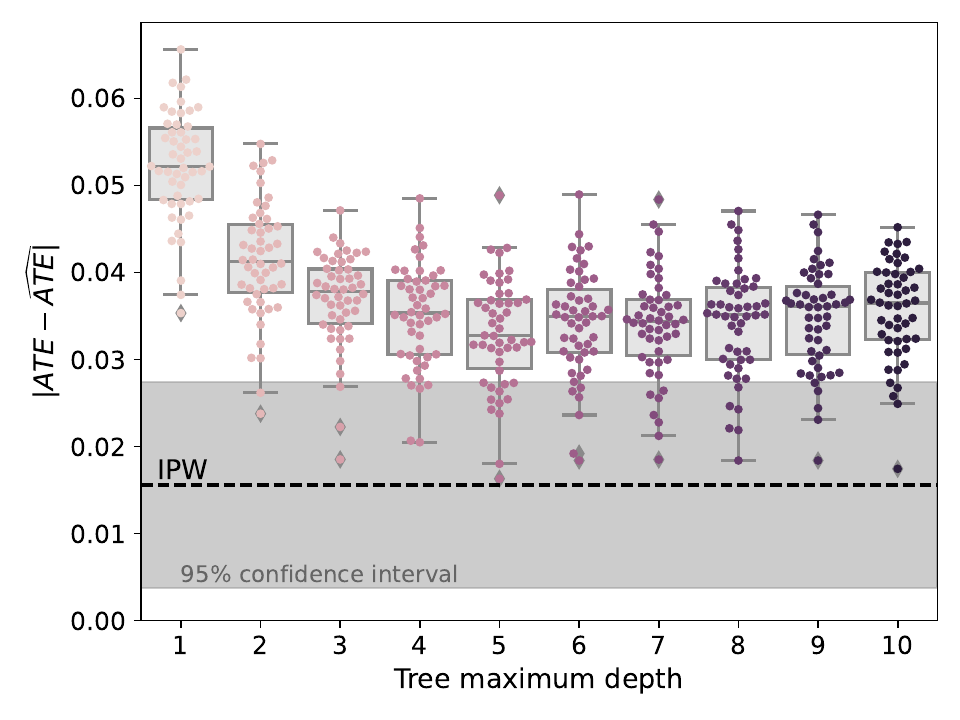}
\caption{Estimation bias for \textit{BICauseTree(Marginal)} with varying maximum depth parameter, and average bias of IPW (dotted), on the twins training set. The estimation bias in leaf nodes decreases for increasing maximum depth values of our tree, and stays consistent for values beyond 5.}
\label{fig:twins_bias_depths}
\end{figure}

We expect a bias-interpretability tradeoff, where deeper trees are less biased but more complex to understand, while shallower trees are less accurate but easier to comprehend.
Figure \ref{fig:twins_bias_depths} shows how estimation bias in leaf nodes decreases as we increase the maximum depth hyperparameter of our BICauseTree(Marginal) in the twins dataset (Figure \ref{fig:bias_depth_acic} for ACIC). Here, each circle represents the estimation bias for a subsample, and the dotted line shows the average bias with an IPW estimator. The shaded area represents the $95\%$  confidence interval (CI) for IPW. The plot shows there is some overlap between the $95\%$ CI for IPW and the estimation bias of deeper trees. The remaining gap thus represents the need for a more complex outcome model in the leaves i.e.\ the estimation bias that was traded against interpretability here.
See how augmenting BICauseTree with an IPW model in the leaves decreases the estimation bias at the cost of transparency. 
Ultimately, figure \ref{fig:twins_bias_depths} shows that bias reduction is consistent beyond a maximum depth parameter of 5. This robustness w.r.t the maximum depth hyperparameter may be due to the pruning step. 

\begin{figure}
\centering
\includegraphics[width=\textwidth]{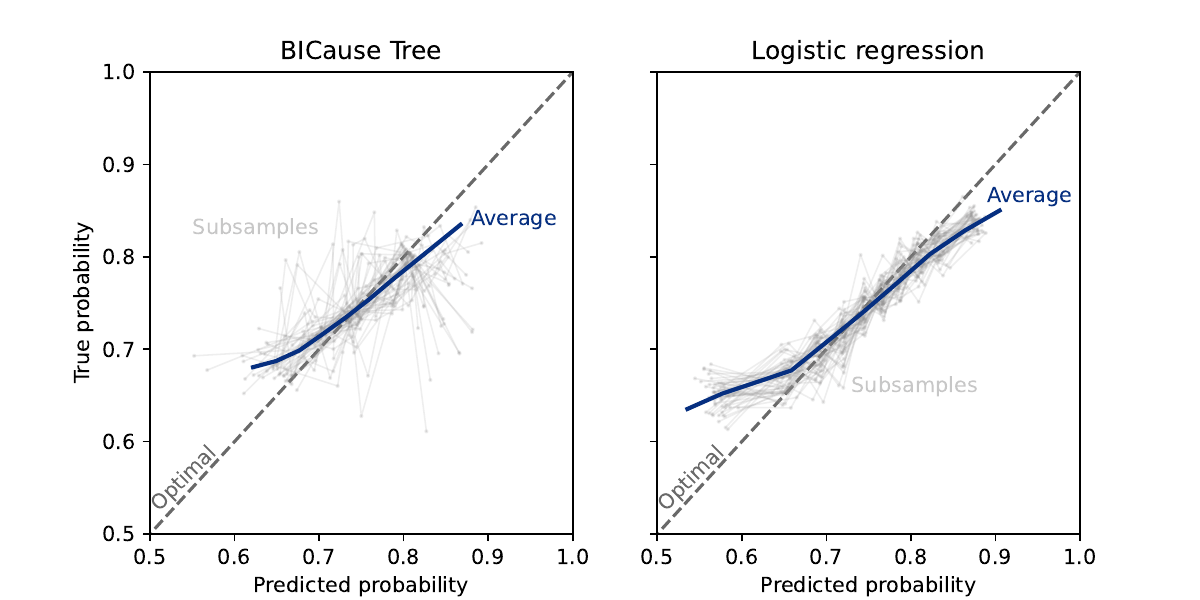}    
\caption{Calibration of propensity score, twins dataset. Though Logsitic regression, which has better data efficiency, has less-noisy calibration, BICauseTree still shows satisfying calibration on average. }
\label{fig:flt_calib_pscore_twins}
\end{figure}

\paragraph{Interpretable positivity violations filtering }\label{sec:interpretability}

As previously discussed, BICauseTree provides a built-in method for identifying positivity violations in the covariate space directly. After positivity filtering, the effect was computed on an average of $99.5\%$ ($\sigma = 0.006$) of the population on the twins dataset, and an average of $85.9\%$ ($\sigma = 0.093$)  of the ACIC dataset. 

Figure \ref{fig:tree_entire_twins} shows the tree partition for the twins dataset. One leaf node was detected as having positivity violations ($N=106$). The twins example illustrates the real-world impact of having a covariate-based definition of the non-violating subpopulation; we are able to claim that our estimated effect of being born heavier might not be valid for newborns that fit the criteria for this specific violating node. This capability of BICauseTree is highly valuable in any safety-sensitive setting. 
Consider a scenario where newborns in that violating subgroup have a higher risk than the rest. Extrapolating the estimated ATE to the entire cohort may therefore prevent them from any additional follow-up visits and thus place them at a higher risk of death.
Knowing who these newborns are would not be possible using an alternative non-interpretable model that provides opaque exclusion criteria, like propensity trimming.
Note that, unlike for effect estimation, the positivity identification remains transparent regardless of the chosen propensity or outcome model at the leaves.

\paragraph{Propensity score estimation}
BICauseTree can also act as a propensity model, modelling the treatment at its leaves.
Given the importance of calibrated propensity scores \cite{gutman2022propensity}, Figure \ref{fig:flt_calib_pscore_twins} compares the calibration of the propensity score estimation of BICauseTree with the one from logistic regression (IPW) on the testing set of the twins dataset (Figure \ref{fig:acic_prop_calibration} for ACIC). As expected, logsitic regression, which has better data efficiency, has better, less-noisy calibration. However, BICauseTree still shows satisfying calibration on average. 
See Appendix for propensity score calibration results on the synthetic natural experiments and the calibration of BICauseTree for the \emph{outcome} variable.

\paragraph{Tree consistency}
To evaluate the consistency of our clustering across subsamples, we train our tree on 70\% of the dataset and compute the adjusted Rand index \cite{rand1971objective} (further details in section \ref{sec:tree_consistency}). We chose not to train on 50\% of the data here as most of the inconsistency would then be due to the variance between subsamples. For the twins dataset, the Rand index across 50 subsamples of size $N=8,388$, is $0.633$ ($\sigma = 0.208$). For the ACIC dataset, the Rand index across 50 subsamples of size $N=3,361$, is $0.314$ ($\sigma = 0.210$) which shows that our tree is not consistent across subsamples if the sample size is not substantial. However, we exemplify consistent identification of the positivity population, with the variance of the percentage of positive samples being $\sigma = 0.006$ and $\sigma = 0.093$ (see paragraph \ref{sec:interpretability}) in the twins and ACIC dataset respectively. Throughout our experiments, we noticed how consistency starts to decrease if the maximum depth hyperparameter increases past a certain threshold, and thus recommend users test the tree consistency across subsamples when tuning this hyperparameter. 

\section{Discussion}\label{sec:discussion}
\vspace{-0.2cm}
We have introduced a decision tree-based model, with a customized splitting criterion, able to estimate the average treatment effect in a transparent, interpretable way.
Additionally, the model can also identify subgroups violating the positivity assumption, characterize them in covariate space, and abstain from estimating an effect on them.
We demonstrated its performance on synthetic data, both toy data and realistic data, allowing us to study the tradeoffs it makes between accuracy and interpretability, and showing it can have comparable accuracy to commonly used methods.

In terms of the bias-interpretability tradeoff, we acknowledge our model may leave more residual confounding bias relative to other less interpretable models in complex data settings.
Nudging the tree to be shallow and more intelligible can force an overly simplistic model of the data that may fail to fully debias the estimated treatment-outcome association.

Nonetheless, this work demonstrated the accuracy and precision of BICauseTree remain reasonably comparable to other common methods, while being far more transparent. Additionally, BICauseTree can detect, characterize, and exclude subgroups violating the overlap assumption. This improves internal validity, as it ensures extrapolation of outcomes across groups is data-based and not model hallucinations. It also improves external validity, informing policy makers to what population---explicitly characterized in covariate space---the estimated effect from the sample is expected to generalize to. On top of being easy to understand, BICauseTree inherits additional desirable strengths from decision trees, like identifying complex interactions in a flexible, data-adaptive, non-parameteric way 
in high-dimensional data (see Figure \ref{fig:high_dim_bias}),
with the computational expense being comparable to IPW and Causal Forest \cite{sani2018computational}.

Our work however has some limitations. First, while our method inherits the inherent interpretability of decision trees, it also inherits their weaknesses. Namely, being a nonparametric estimator, it has lower data efficiency than other parametric estimators, like the common IPW with a logistic regression estimator. Theoretically, this should result in higher variance, but since IPW is a low-precision model, our experiments show their variance is comparable.

Additionally, decision trees use fewer samples as they grow, decreasing the precision of the estimated splitting criterion. 
Therefore, while the mean of the estimated $\widehat{\text{ASMD}}$, may be independent of sample size, its variance is dependent and can become less precise \cite{austin2009balance}.
This holds true to our ASMD-based splitting criterion. 
Furthermore, splitting covariates solely depend on ASMD, which, is only a marginal measure of inbalance, and does not regard the association of the covariate with the outcome variable 
(see \ref{sec:asmd_justification}). 

Future work could perfect some of the limitations presented above.
First, the pruning procedure correcting for multiple hypotheses can avoid assuming independence and leverage the hierarchical tree structure \cite{behr2020testing, allen1997testing, malek2017sequential}. 
Second, following the ``honest effect'' framework \cite{athey2016recursive} we can further split the sample, using one split to structure the tree and another to fit the models at its nodes; though this will require even more data. 
Third, we currently fit models at each leaf independently, but since our target estimand is the ATE, we can partially pool estimates across leaves (clusters), fitting a multilevel outcome model with varying intercepts or varying slopes for treatment coefficients \cite{feller2015hierarchical}. 
Lastly, future work may further explore additionally excluding terminal leaf nodes with high imbalance (as they might not enclose natural experiments), or experiment with partially applying additional covariate adjustments in those unbalanced leaves, while balanced use a simple, interpretable average for estimation.

Limitations notwithstanding, we claim BICauseTree may be relevant when causation is examined in high-stake or public-facing domains. Its transparency can inspire trust among laypeople, while any concerns about its accuracy may be mitigated by fitting an ensemble of causal models and examining if the BICauseTree estimator is an outlier.
We provide an open-source implementation of the model in Python adhering to \texttt{causallib}'s API for maximizing ease of use and encouraging practitioners to use it in their own work.

\newpage

\bibliography{bib}
\bibliographystyle{ieeetr}

\newpage


\onecolumn
\appendix

\section{Appendix}

\renewcommand{\thefigure}{A\arabic{figure}}
\setcounter{figure}{0}
\renewcommand{\thetable}{A\arabic{table}}
\setcounter{table}{0}

\subsection{Causal inference}\label{sec:causal_inf_app}

\paragraph{Confounder}{A confounder is a variable that is associated with both the treatment and the outcome, causing a spurious correlation. For instance, summer is associated with eating ice cream and getting sunburns, but there is no causal relationship between the two.}
\paragraph{Propensity score model}{A propensity score model is a function that predicts treatment from the observed covariates i.e.\ $P(T=1|C=c)$ for a binary treatment $T$ and a covariate vector $C$.}
\paragraph{Potential outcome}{As defined by the Rubin causal model \cite{rubin2005causal}, a potential outcome $Y(t)$ is the value that $Y$ would take if $T$ were set by (hypothetical) intervention to the value $t$.}
\paragraph{Identification assumptions}
Inference is possible under three identification assumptions. 
\begin{itemize}
    \item \textbf{No interference} For a given individual $i$, this assumption implies that $Y_i(t)$ represents the value that $Y$ would have taken for individual i if $T$ had been set to $t$ for individual $i$, i.e\ the potential value of $Y_i$ if $T_i$ had been set to $t$.
    \item \textbf{Consistency} For a given individual $i$, $T_{i}=t \Rightarrow Y_{i}=Y_{i}(t)$. This means that for individuals who actually received treatment level $t$, their observed outcome is the same as what it would have been had they received treatment level $t$ via a hypothetical intervention. 
    \item \textbf{Conditional exchangeability} For a given individual $i$, we assume that conditional on $C$, the actual treatment level $T$ is independent of each of the potential outcomes: \\
    $Y(t) \perp T \mid \mathbf{C}, \forall t$
\end{itemize}

\subsection{Evaluation of the tree consistency}\label{sec:tree_consistency}
We evaluated the consistency of our clustering across subsamples using an Adjusted Rand Index \cite{rand1971objective}. Given a set of $n$ elements $S=\left\{o_1, \ldots, o_n\right\}$ and two partitions of $S$ to compare, $X=\left\{X_1, \ldots, X_r\right\}$, a partition of $S$ into $r$ subsets, and $Y=\left\{Y_1, \ldots, Y_s\right\}$, a partition of $S$ into $s$ subsets, define the following:
\begin{itemize}
    \item $a$, the number of pairs of elements in $S$ that are in the same subset in $X$ and in the same subset in $Y$.
    \item $b$, the number of pairs of elements in $S$ that are in different subsets in $X$ and in different subsets in $Y$.
    \item $c$, the number of pairs of elements in $S$ that are in the same subset in $X$ and in different subsets in $Y$.
    \item $d$, the number of pairs of elements in $S$ that are in different subsets in $X$ and in the same subset in $Y$.
\end{itemize}

The Rand index, $R$, is: 
$$
R=\frac{a+b}{a+b+c+d}=\frac{a+b}{\left(\begin{array}{l}
n \\
2
\end{array}\right)}
$$
Intuitively, $a+b$ can be considered as the number of agreements between $X$ and $Y$ and $c+d$ as the number of disagreements between $X$ and $Y$. The adjusted Rand index is the corrected-for-chance version of the Rand index \cite{hubert1985comparing}.

\subsection{Further related work}
\label{sec:further_related_work}
Causal inference provides a wide range of methods for estimating causal effect from data with unbalanced treatment allocation. In balancing methods such as matching or weighting methods, the data is pre-processed to create subgroups with lower treatment imbalance or  ``natural experiments''.

\paragraph{Matching}

\emph{Matching} methods consist of clustering similar units from the treatment and control groups to reduce imbalance. In general, a matching procedure generates weights $w_{i j}$ denoting the assignment of one or many control units $j$ to a treated unit $i$ (\cite{morgan2015counterfactuals}, Chapter 5). Exact matching only assigns control units to treatment units with the exact same set of covariate values. But typically, matched control units $j$ are chosen based on a nearest-neighbors search according to some distance metric. However, matching procedures induce a bias-variance trade-off as discarding unmatched samples reduces estimation error at the cost of increased variance. However, all matching methods suffer from the curse of dimensionality, making them impractical in high-dimensional datasets. Exact matching and coarsened exact matching \cite{iacus2012causal} find exponentially fewer matches as the input dimension grows \cite{abadie2006large}. Alternative methods include Propensity score matching \cite{austin2011optimal}, where distance is computed from an estimate of the propensity score $P(T=1 \mid X=\boldsymbol{x})$, and Mahalanobis distance matching \cite{rubin1980bias} (see more details below in \ref{paragraph:baseline_comp}). However, compression into a single dimension can lead to highly unrelated matches with very different characteristics in the original covariate space, and can ultimately increase estimation bias \cite{king2019propensity}. \textit{Clivio et. al} overcome this issue by developing a multivariate balancing score to perform matching for high-dimensional causal inference. Nevertheless, this approach is not interpretable.

\paragraph{Weighting methods}

An alternative to matching are \emph{weighting} methods, where sample weights are estimated, generalizing the problem formulation of matching. In Inverse Probability Weighting (IPW) \cite{horvitz1952generalization} which is the most popular alternative in that category, the samples are weighted according to their \emph{propensity} score, i.e.\ the estimated probability of treatment conditional on their covariates.

\paragraph{Adjustment methods}

\emph{Adjustment} methods estimate the causal effect from regression outcome models where both treatment and covariates act as predictors of the outcome. These regressions can be fitted through various methods like linear regression \cite{imbens2015causal}, neural networks \cite{shi2019adapting, shalit2017estimating}, or tree-based models \cite{athey2016recursive}. Common alternatives include Doubly Robust estimators \cite{seaman2018introduction}, Double Debiased Machine Learning \cite{chernozhukov2018double} and metalearners such as the T-learner and X-learner \cite{kunzel2019metalearners}. These methods have the advantage of being very data-adaptive. 
More particularly, the Causal Tree approach \cite{athey2016recursive} builds on regression tree methods, and splits the data to optimize for goodness of fit in treatment effects. Causal Tree separates the training dataset into two subsamples: a splitting subsample and an estimating subsample. The splitting subsample is used to build a causal tree while the estimating subsample is used to generate unbiased conditional treatment effect estimates. This procedure is called ``honest estimation'' and is anticipated to avoid overfitting.

\subsection{Experimental details: synthetic datasets}\label{sec:exp_details}

\subsubsection{Natural experiment dataset}\label{sec:natural_exp}

For the natural experiment dataset, we consider a Death outcome $D$, a binary treatment of interest $T$ and two covariates such that $X = (S, A)$ with $S$ the sex and $A$ the continuous age such that: \\
$$ S \sim \operatorname{Bernoulli}(0.5)$$
$$ A\sim \operatorname{Normal}\left(50 , 20^2\right)$$ 
The sample size was chosen to be $N=20,000$.

We defined four sub-populations, each constituting a natural experiment, with a different propensity distribution $P(T=1 | X=x)$: 

$$\Pr( T=1 | S=1, A \geq 50) \sim  \operatorname{TruncatedNormal}_{[0,1]}\left(0.5 , 0.1^2\right)$$
$$\Pr( T=1 | S=1, A < 50) \sim  \operatorname{TruncatedNormal}_{[0,1]}\left(0.3 , 0.1^2\right)$$
$$\Pr( T=1 | S=0, A \geq 50) \sim  \operatorname{TruncatedNormal}_{[0,1]}\left(0.1 , 0.1^2\right)$$
$$\Pr( T=1| S=0, A < 50) \sim  \operatorname{TruncatedNormal}_{[0,1]}\left(0.4 , 0.1^2\right)$$

Individual treatment propensities were sampled from the corresponding distributions above and observed treatment values were sampled from a Bernoulli distribution parameterized with the individual propensities. 
The outcome probabilities were not modeled as a distribution. Instead, observed outcome values were sampled directly from a Bernoulli distribution parameterized with a constant value that depended on both $X$ and $T$. 
\begin{itemize}[noitemsep]
    \item $\Pr(Y | T=1, S=1, A \geq 50) \sim \mathcal{B}(0.1)$ 
    \item $\Pr(Y | T=1, S=1, A < 50) \sim \mathcal{B}(0.2)$ 
    \item $\Pr(Y | T=1, S=0, A \geq 50) \sim \mathcal{B}(0.4)$ 
    \item $\Pr(Y | T=1, S=0, A < 50) \sim \mathcal{B}(0.15)$ 
    \item $\Pr(Y | T=0, S=1, A \geq 50) \sim \mathcal{B}(0.2)$ 
    \item $\Pr(Y | T=0, S=1, A < 50) \sim \mathcal{B}(0.4)$ 
    \item $\Pr(Y | T=0, S=0, A \geq 50) \sim \mathcal{B}(0.8)$ 
    \item $\Pr(Y | T=0, S=0, A < 50) \sim \mathcal{B}(0.3)$ 
\end{itemize}
No positivity violation was modeled in this experiment.

\subsubsection{Positivity violations dataset}\label{sec:positivity}

For this second synthetic experiment, we build a dataset with two positivity-violating subgroups. Let us consider a synthetic example of a dataset with a Death outcome $D$, a binary treatment of interest $T$, and three binary covariates --sex, cancer, and arrhythmia-- such that $X=(S, C, A)$. For the marginal distributions, we set: \\
$$ S \sim \operatorname{Ber}(0.5)$$
$$ C \sim \operatorname{Ber}(0.3)$$
$$ A \sim \operatorname{Ber}(0.1)$$
The sample size was chosen to be $N=20,000$.
\begin{itemize}[noitemsep]
    \item $\Pr( T=1 | S=1, C=1, A=1) \sim \operatorname{TruncatedNormal}_{[0,1]}\left(1.00 , 0.02^2\right)$
    \item $\Pr( T=1 | S=1, C=0, A=1) \sim \operatorname{TruncatedNormal}_{[0,1]}\left(0.32 , 0.10^2\right)$
    \item $\Pr( T=1 | S=1, C=1, A=0) \sim \operatorname{TruncatedNormal}_{[0,1]}\left(0.12 , 0.10^2\right)$ 
    \item $\Pr( T=1 | S=1, C=0, A=0) \sim \operatorname{TruncatedNormal}_{[0,1]}\left(0.42 , 0.10^2\right)$
    \item $\Pr( T=1 | S=0, C=1, A=0) \sim \operatorname{TruncatedNormal}_{[0,1]}\left(0.17 , 0.10^2\right)$
    \item $\Pr( T=1 | S=0, C=1, A=1) \sim \operatorname{TruncatedNormal}_{[0,1]}\left(0.30 , 0.10^2\right)$
    \item $\Pr( T=1 | S=0, C=0, A=1) \sim \operatorname{TruncatedNormal}_{[0,1]}\left(0.24 , 0.10^2\right)$
    \item $\Pr( T=1 | S=0, C=0, A=0) \sim \operatorname{TruncatedNormal}_{[0,1]}\left(0.00, 0.02\right)$
\end{itemize}

The observed outcome values were sampled from a Bernoulli distribution parameterized with a constant value that depended on both $X$ and $T$. 
For the treated:
\begin{itemize}[noitemsep]
    \item $\Pr(Y | T=1, S=1, C=1, A=1) \sim \mathcal{B}(0.13)$ 
    \item $\Pr(Y | T=1, S=1, C=1, A=0) \sim \mathcal{B}(0.08)$ 
    \item $\Pr(Y | T=1, S=1, C=0, A=1) \sim \mathcal{B}(0.21)$ 
    \item $\Pr(Y | T=1, S=1, C=0, A=0) \sim \mathcal{B}(0.1)$ 
    \item $\Pr(Y | T=1, S=0, C=1, A=1) \sim \mathcal{B}(0.36)$ 
    \item $\Pr(Y | T=1, S=0, C=1, A=0) \sim \mathcal{B}(0.29)$ 
    \item $\Pr(Y | T=1, S=0, C=0, A=1) \sim \mathcal{B}(0.24)$ 
    \item $\Pr(Y | T=1, S=0, C=0, A=0) \sim \mathcal{B}(0.09)$ 
\end{itemize}
For the untreated:
\begin{itemize}[noitemsep]
    \item $\Pr(Y | T=0, S=1, C=1, A=1) \sim \mathcal{B}(0.31)$ 
    \item $\Pr(Y | T=0, S=1, C=1, A=0) \sim \mathcal{B}(0.4)$ 
    \item $\Pr(Y | T=0, S=1, C=0, A=1) \sim \mathcal{B}(0.29)$ 
    \item $\Pr(Y | T=0, S=1, C=0, A=0) \sim \mathcal{B}(0.45)$ 
    \item $\Pr(Y | T=0, S=0, C=1, A=1) \sim \mathcal{B}(0.4)$ 
    \item $\Pr(Y | T=0, S=0, C=1, A=0) \sim \mathcal{B}(0.51)$ 
    \item $\Pr(Y | T=0, S=0, C=0, A=1) \sim \mathcal{B}(0.43)$ 
    \item $\Pr(Y | T=0, S=0, C=0, A=0) \sim \mathcal{B}(0.73)$ 
\end{itemize}


\clearpage
\newpage

\subsection{Further experiment results: synthetic experiments}\label{sec:further_sim}

\subsubsection{Natural experiment dataset}\label{sec:further_sim1}

\begin{figure}[H]
    \centering
    \includegraphics[width=\textwidth]{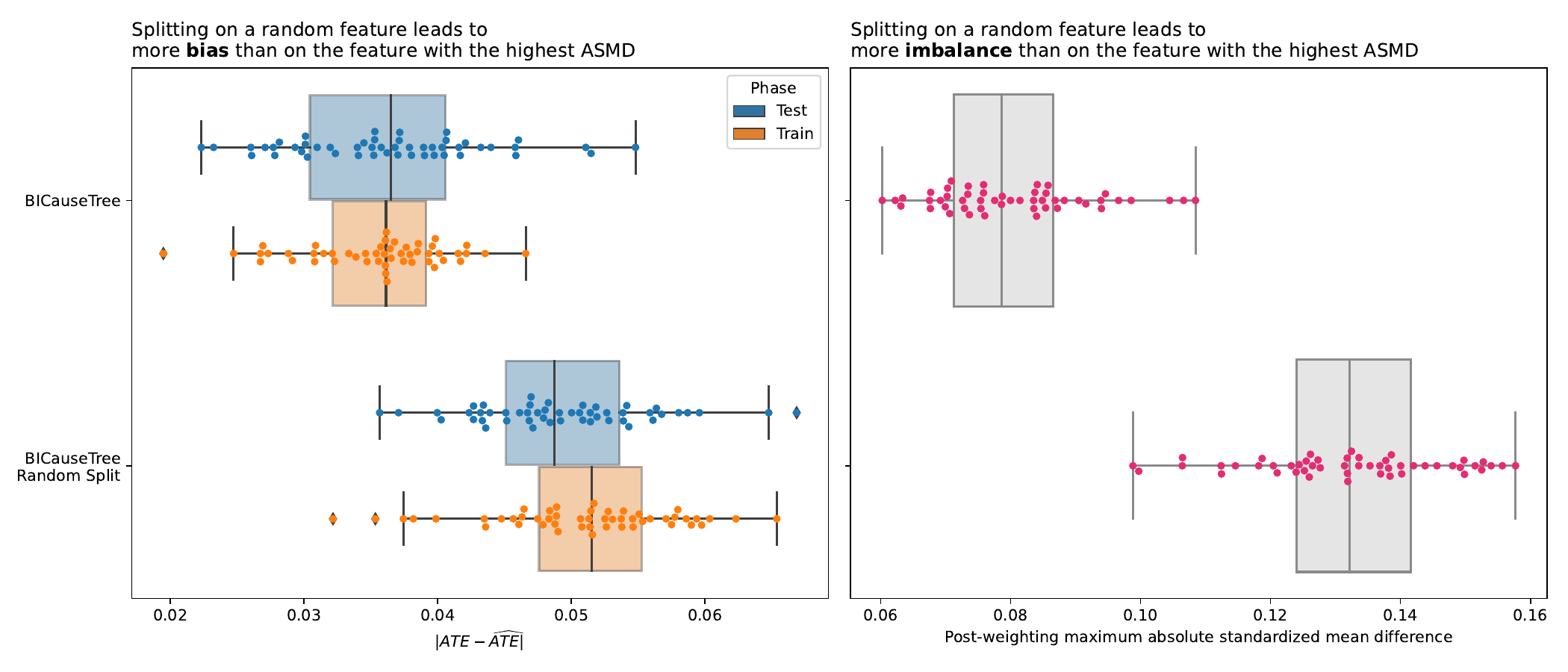}
    \caption{Comparison of estimation bias (left, absolute difference) and imbalance (right, maximum absolute standardized mean difference [ASMD]) using two flavors of BICauseTree: First, the original version from the manuscript, wherein each recursion the feature to split on is chosen by the one with maximal ASMD. And second, a variant where features are selected at random, with all other hyperparameters held constant. We see that selecting the most imbalanced feature for stratification leads to better balancing and, more importantly, better estimation - justifying the rationale for selecting features based on the highest ASMD.}
    \label{fig:random_split}
\end{figure}

\paragraph{The importance of selecting candidate features by max ASMD} \label{sec:asmd_justification}
The heuristic for choosing the feature with the maximum absolute standardized mean difference (ASMD) among all features to split on can be empirically justified.
We can do that by comparing two BICauseTrees: one that selects the feature with maximal ASMD at each recursive iteration (the model used throughout this work),
to a second tree whose only difference is choosing candidate features randomly. 
Figure \ref{fig:random_split} demonstrates how this design decision contributes to a reduction in estimation bias.

Additionally, adjusting for the maximal ASMD can be justified in simplistic scenarios.
Consider the following data generating process:
\begin{equation*}
\begin{split}
    Y &= A + X_1 + X_2 \\
    A &\sim Bernoulli(0.5)  \\
    X_1, X_2 &\sim Normal(0,1)  \\
    X_2 |_{A=1} &\sim Normal(1,1)  \\
\end{split}
\end{equation*}

Where $X_1$ is balanced (independent of $A$), but $X_2$ is not since the mean of the treated samples is shifted one unit upward.

\begin{figure}[H]
    \centering
    \includegraphics[width=0.9\textwidth]{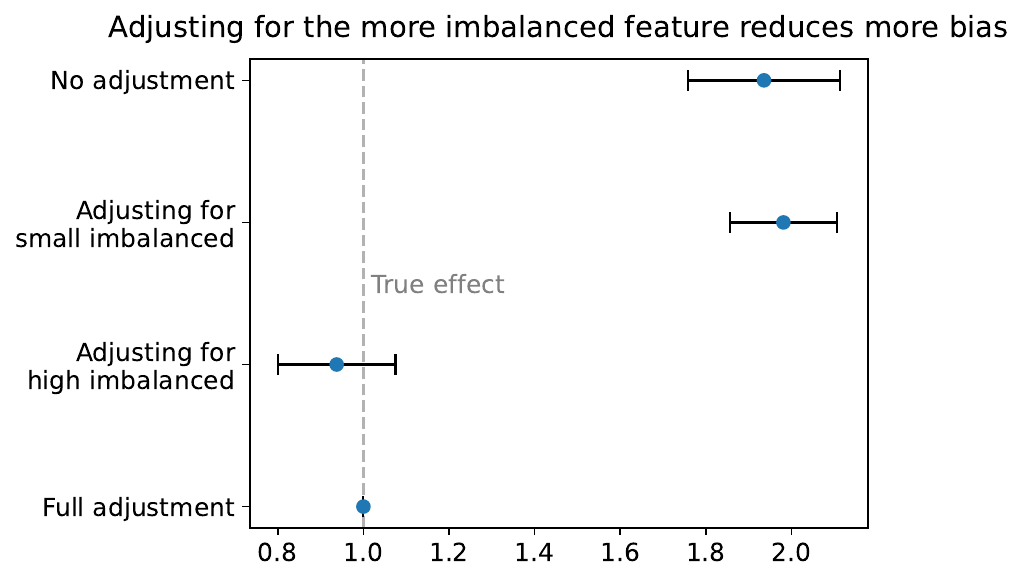}
    \caption{Adjusting for the more imbalanced predictor reduces more estimation bias of the treatment effect. 
    The data contains a treatment assignment and two predictor variables -- one balanced across treatment groups and one imbalanced. 
    We examine four models, all adjusting for the treatment assignment but differing in their adjustment of the predictors. 
    Adjusting for the more imbalanced predictor almost retrieves the treatment effect perfectly (dashed grey line), 
    while adjusting the less imbalanced predictor still suffers from large residual confounding.}
    \label{fig:asmd_adjustment}
\end{figure}

We can then consider four simple linear models: \\
1) Null-model: $Y \sim 1 + A$. \\ 
2) Full-model: $Y \sim 1 + A + X_1 + X_2$. \\ 
3) $X_1$-model: $Y \sim 1 + A + X_1$. \\ 
4) $X_2$-model: $Y \sim 1 + A + X_2$. \\
And observe that the $X_2$-model recovers the true effect much better than the $X_1$-model (see Figure \ref{fig:asmd_adjustment}).

Note that this implicitly assumes all predictors have an equal association with the outcome.
Therefore, as suggested in the main text, future work can create a combine score of ASMD and predictor-outcome association.
One such approach can be as follows.
\begin{enumerate}
    \item We calculate the ASMDs, denoted $s$. 
    \item We then calculate any variable importance measure, preferably fitted jointly using all predictors and treatment assignment, as long as its values ranges in the positives. Denote it as $q$.\\
    Examples can be the absolute coefficients for regression models or permutation importance for tree-based models.
    \item We then normalize both $s$ and $q$, separately so each sum to 1, depicting relative imbalance and relative predictor-outcome importance among predictors.
    \item Finally, we use the combined measurement $s \cdot q$ to select the predictor to split on in each iteration. 
\end{enumerate}

Examining $s \cdot q$ it is easy to see that if a variable $j$ is balanced (near-zero $s_j$) or unrelated to the outcome (near-zero $q_j$), the combined score will be small and less likely to be chosen for splitting. 
The predictor $j$ itself is also less likely to cause bias because it is either balanced, or, if imbalanced, affects the outcome very little.

\begin{figure}[H]
    \centering
    \includegraphics[width=\textwidth]{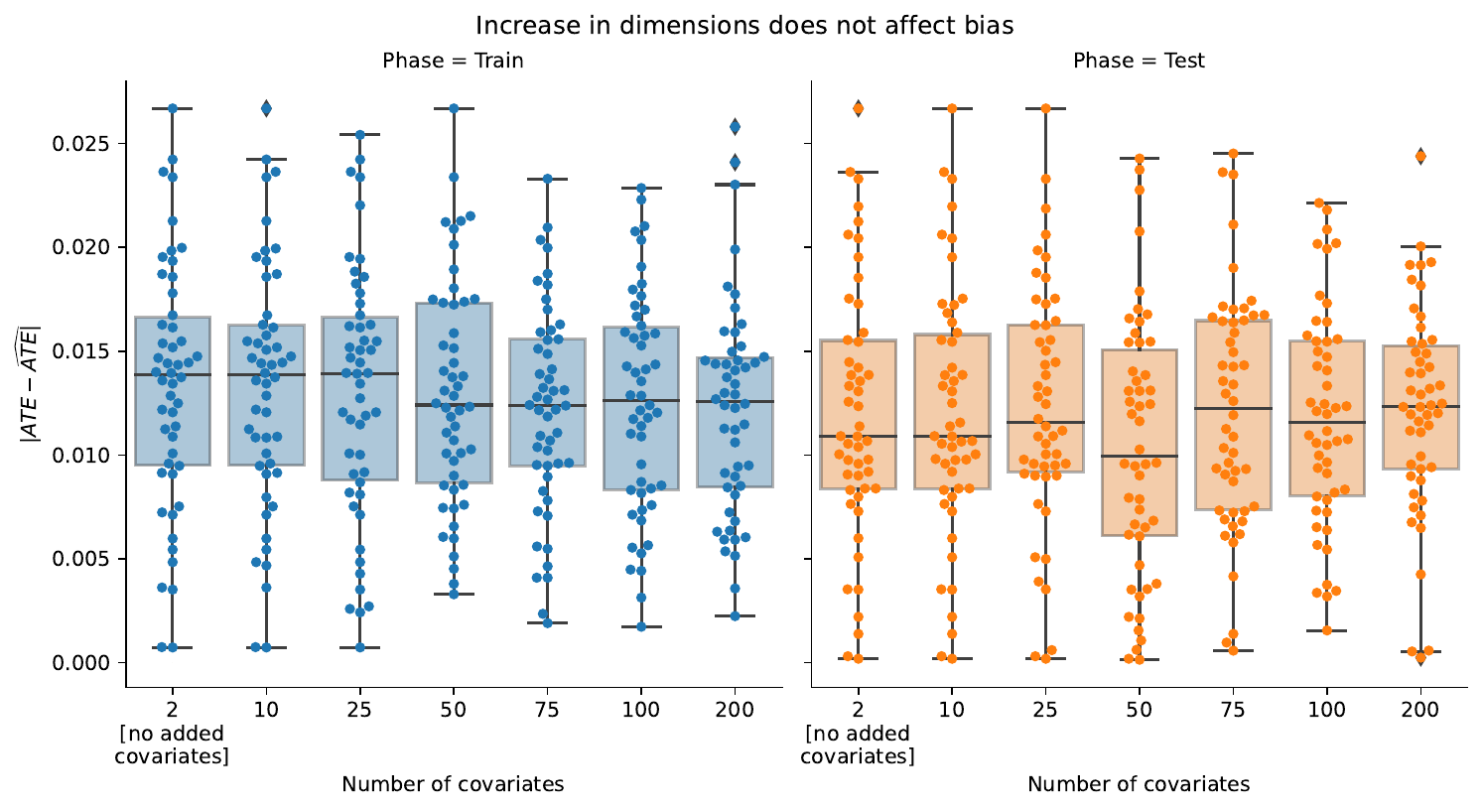}
    \caption{Comparison of estimation bias as a function of number of covariates. We take a synthetic dataset that adheres to conditional exchangeability and increasingly add noisy covariates to it. We see the estimation bias remains the same, throughout the train set (left) and the test set (right) regardless of the number of covariates added.
    This better strengthens our claim for the suitability of our method to discover unbiased natural experiments in high-dimensional data.}
    \label{fig:high_dim_bias}
\end{figure}

\paragraph{Applicability in high-dimensional data}
Being based on decision tree logic allows BICauseTree to scale properly for high-dimensional data. 
BICauseTree, like decision trees, scans all the features before deciding which to split on. 
This allows it to focus on the most biasing feature at each partition, regardless of their number, enabling them to find natural experiments in noisy high-dimensional data.
To demonstrate that, we augmented our natural experiment dataset with an increasing number of noisy features. 
Figure \ref{fig:high_dim_bias} shows that the error of BICauseTree's ATE estimation stays consistent regardless of the number of added features.

\paragraph{Partitioning}
The tree partitions so to recreate the four intended sub-populations where a natural experiment was simulated. The average Rand index was equal to 0.901 across the 50 subsamples ($\sigma=0.263$). Violating leaf nodes are marked in red.
\begin{figure}[H]
    \centering
    \includegraphics[width=\textwidth]{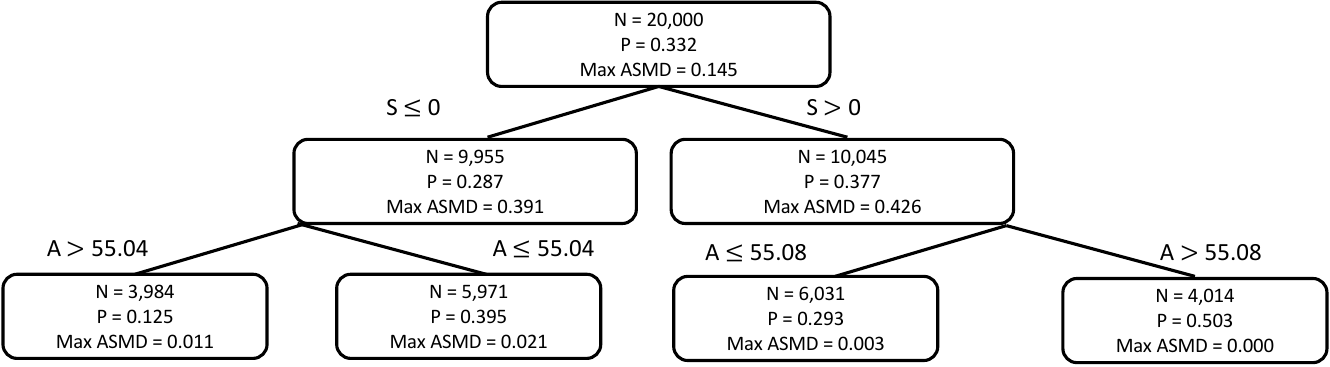}
    \caption{Tree structure after training on the entire natural experiment dataset ($N = 20,000$). Violating leaf nodes are marked in red. P is for node prevalence.}
    \label{fig:tree_entire_sim1}
\end{figure}

\paragraph{Causal effect estimation}\label{sec:further_sim1_effect}
Here, Causal Tree has both higher estimation bias and higher estimation variance compared to other methods. The tree-like nature of our data structure may be incompatible with the optimization function of Causal Tree, which maximizes treatment effect heterogeneity. Causal Forest, which has multiple Causal Tree models, however, overcomes this difficulty and performs well.
\begin{figure}[H]
    \centering
    \includegraphics[width=.5\textwidth]{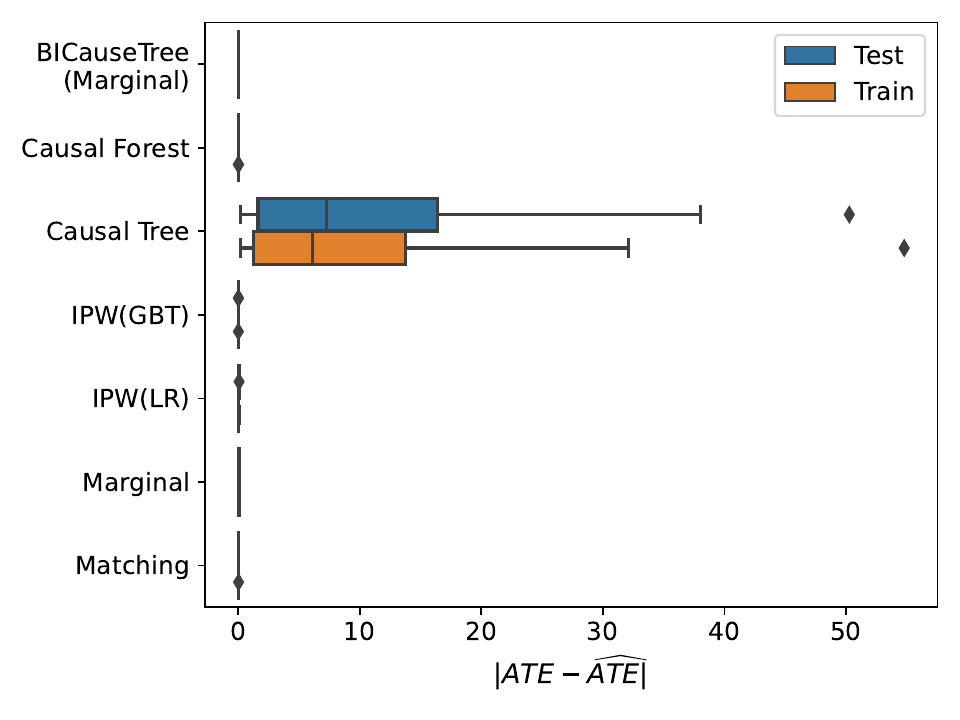}
    \caption{Estimation bias for the natural experiment dataset across 50 subsamples with $N = 20,000$.}
\end{figure}

\paragraph{Propensity score estimation}
BICauseTree(Marginal)'s propensity estimation is well calibrated. It is closer to the identity line than logistic regression- (IPW(LR)) and the gradient boosting trees- (IPW(GBT)) based models. The calibration however remains satisfactory across all models. This further shows our ability to identify natural experiments in a simple data setting. 
\begin{figure}[H]
    \centering
    \includegraphics[width=\textwidth]{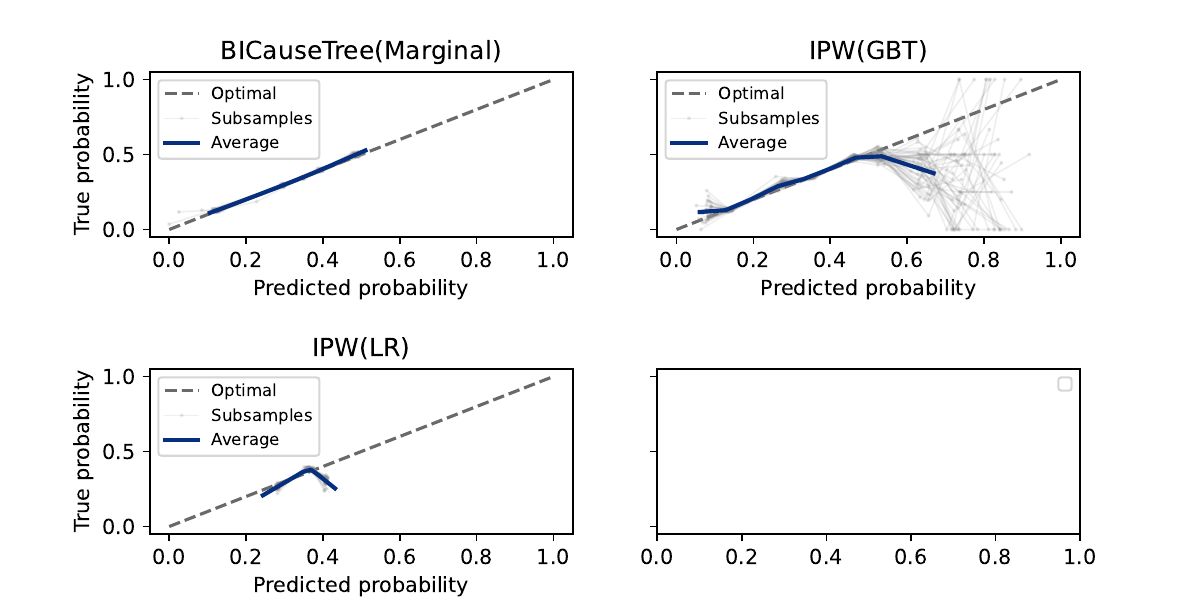}
    \caption{Calibration of the propensity score estimation for the natural experiment dataset across 50 subsamples, on the natural experiment testing set ($N=10,000$).}
\end{figure}

\paragraph{Outcome estimation}
The performance of BICauseTree w.r.t outcome estimation is only satisfactory in this experiment. This shows our ability to estimate the ATE despite a somehow lower calibration of our outcome estimation. Ultimately, the performance would likely have been improved had we used a more complex outcome model. The calibration of the predicted outcomes by BICauseTree however remains better than the one by Matching. 

\begin{figure}[H]
    \centering
    \includegraphics[width=\textwidth]{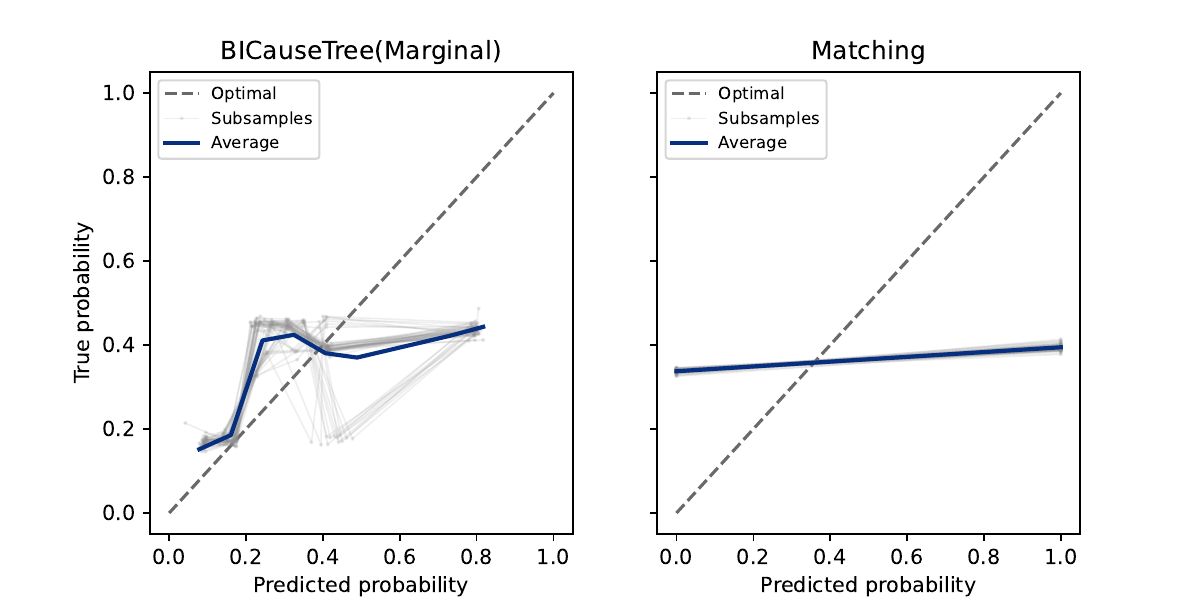}
    \caption{Calibration of the outcome estimation for the natural experiment dataset across 50 subsamples on the natural experiment testing set ($N=10,000$).}
\end{figure}

\begin{figure}[H] 
    \centering
    \includegraphics[width=0.5\textwidth]{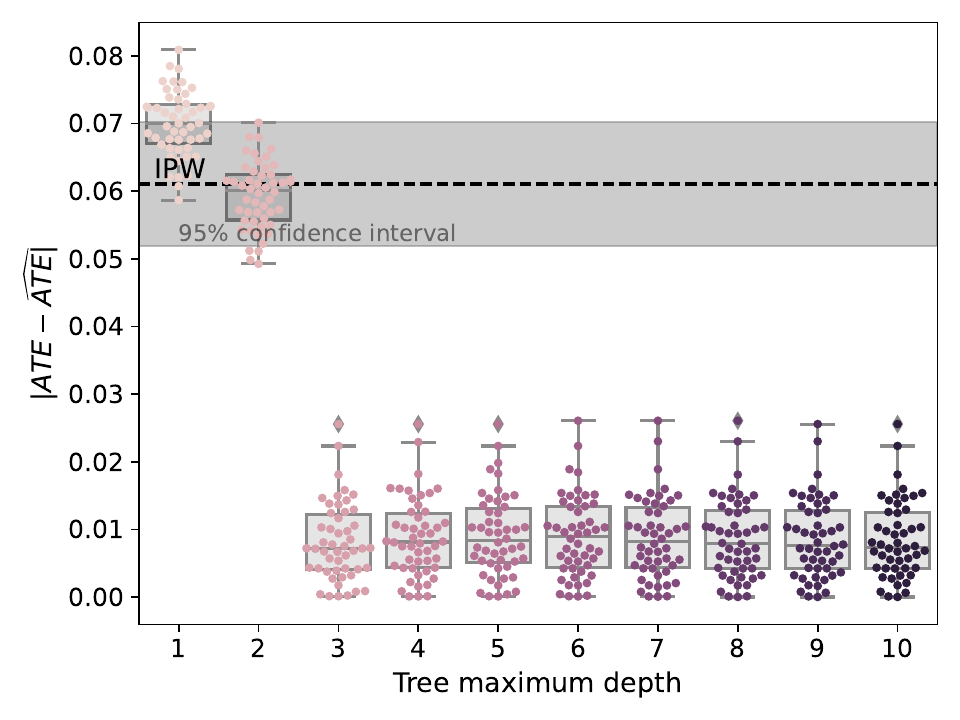}
    \caption{Estimation bias when comparing \textit{BICauseTree(Marginal)} with varying maximum depth parameters with the average bias of IPW (dotted), on the natural experiment training set ($N=10,000$) across 50 subsamples.}
    \label{fig:effect_depth_natural}
\end{figure}

\paragraph{Effect estimation bias reduction with tree depth}
Figure \ref{fig:effect_depth_natural} shows how estimation bias decreases as we increase the maximum depth hyperparameter of our BICauseTree(Marginal) on the natural experiment training dataset ($N=10,000)$. Here, each circle in the plot represents the estimation bias for a subsample. The dotted line shows the average bias with an IPW estimator. The shaded area represents the $95\%$  confidence interval (CI) for IPW. This result shows that our estimation is robust to the choice of a maximum tree depth hyperparameter. Above a certain threshold, the effect estimate from BICauseTree remains unbiased.

\paragraph{Treatment allocation bias reduction with tree depth}
Figure \ref{fig:asmd_natural} below shows the weighted ASMD of both covariates $S$ and $A$ in BICauseTree models with varying maximum tree depth hyperparameters. It illustrates the reduction of treatment allocation bias as the tree grows, but also shows that this reduction is a heuristic as the ASMD might not decrease monotonically.
\begin{figure}[H]
    \centering
    \includegraphics[width=0.5\textwidth]{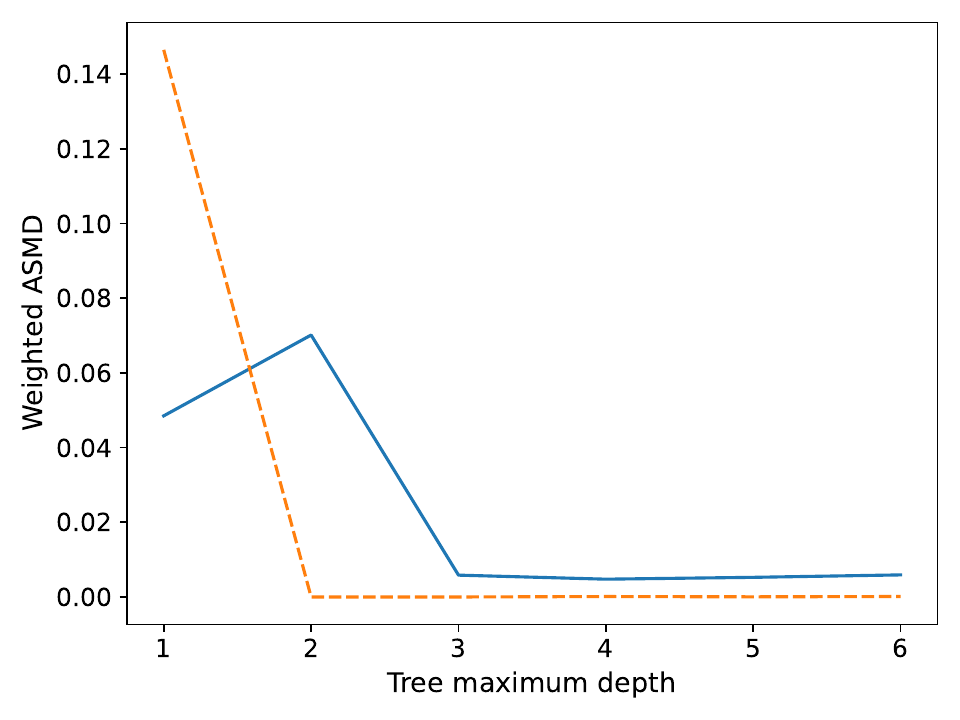}
    \caption{Weighted ASMD for all covariates applying \textit{BICauseTree(Marginal)} models with varying maximum tree depths on the natural experiment training set ($N=10,000$) across 50 subsamples.}
    \label{fig:asmd_natural}
\end{figure}

\subsubsection{Positivity violations dataset}\label{sec:further_sim2}

\paragraph{Partitioning}

BICauseTree coherently flags the two subpopulations where positivity violations were simulated, as shown in the example partition on the entire dataset below in Figure \ref{fig:partition_positivity} (violating leaves are marked in red). 
The average Rand index was equal to 0.986 across the 50 subsamples ($\sigma=0.026$).

\begin{figure}[H]
    \centering
    \includegraphics[width=\textwidth]{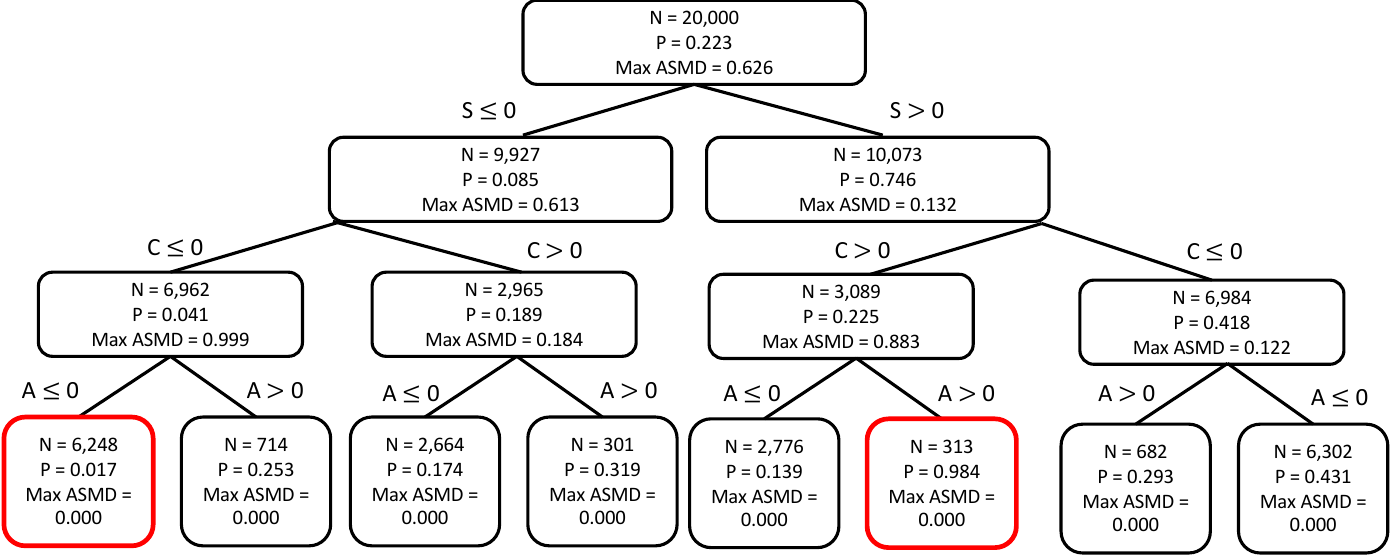}
    \caption{Tree structure after training on the entire positivity violations dataset ($N = 20,000$). Violating leaf nodes are marked in red. P is for node prevalence.}
    \label{fig:partition_positivity}
\end{figure}

\paragraph{Causal effect estimation}\label{sec:further_sim2_effect}

Our tree compares with all existing alternatives. Matching has both higher estimation bias and higher estimation variance compared to other methods. 

\begin{figure}[H]
    \centering
    \includegraphics[width=0.5\textwidth]{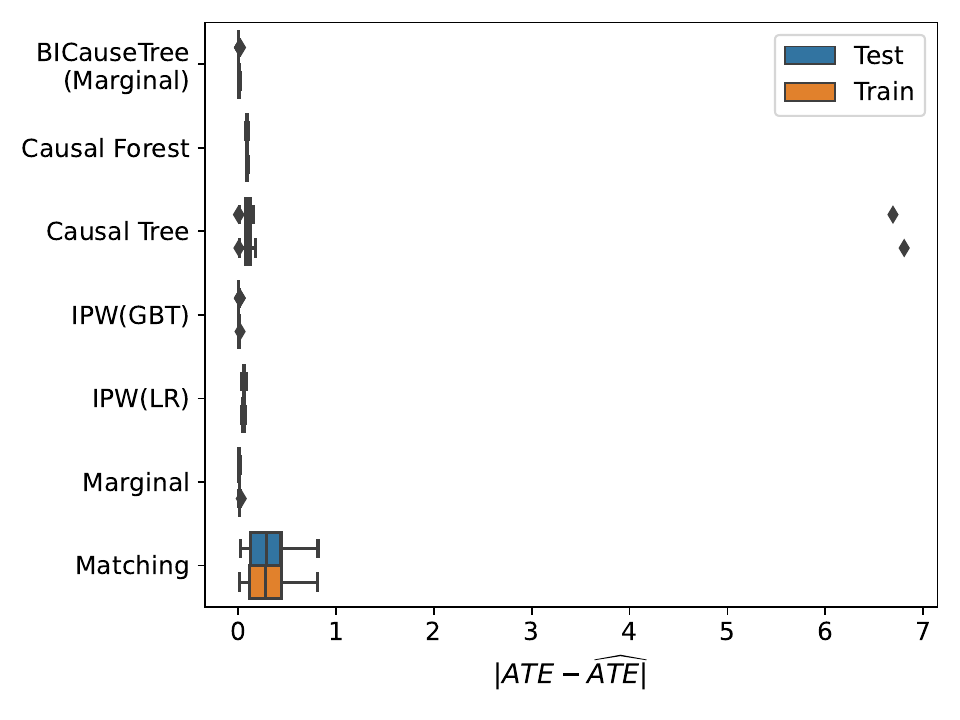}
    \caption{Estimation bias on the positivity violations dataset ($N = 20,000$) across 50 subsamples}
    \label{fig:diff_all_positivity}
\end{figure}

\paragraph{Propensity score estimation}
BICauseTree(Marginal) shows good calibration since it indeed clearly identified the correct subpopulations. 
The gradient-boosting tree (IPW(GBT)) also shows good calibration, as the tree-based modeling captures the underlying structure of the data.
On the other hand, the logistic regression-based model (IPW(LR)) is ill-specified for the data and therefore presents relatively poorer calibration.
\begin{figure}[H]
    \centering
    \includegraphics[width=\textwidth]{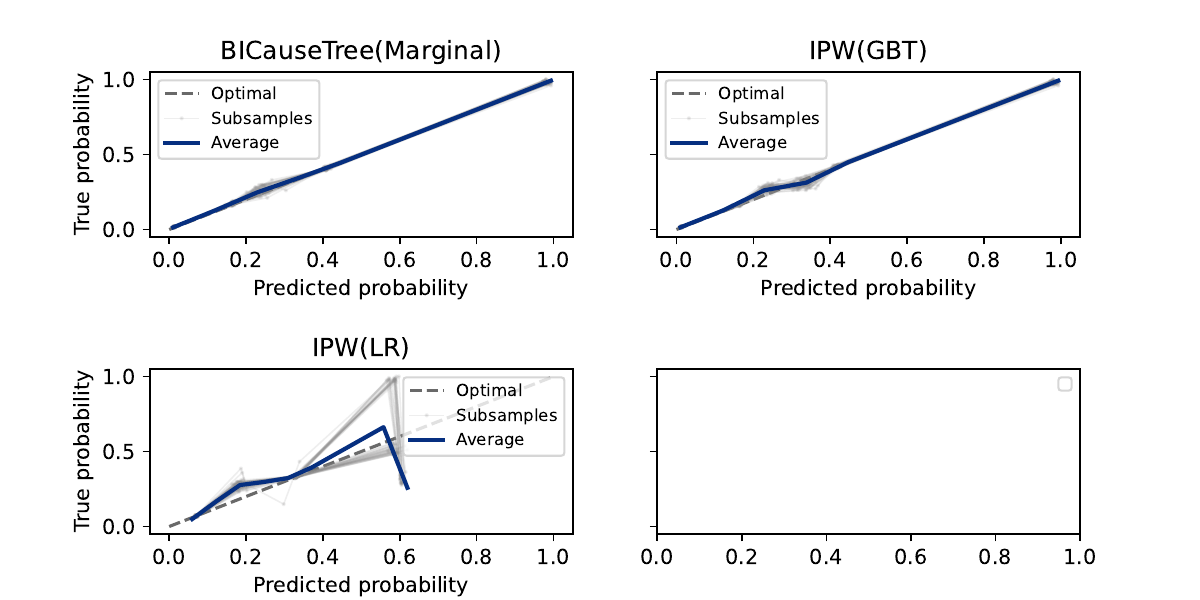}
    \caption{Propensity score calibration comparing our approach to existing alternatives on the testing set of the positivity violation dataset ($N = 10, 000$) across 50 subsamples}
\end{figure}

\paragraph{Outcome estimation}
BICauseTree shows descent calibration on outcome prediction in this experiment. Matching shows poor calibration. 

\begin{figure}[H]
    \centering
    \includegraphics[width=\textwidth]{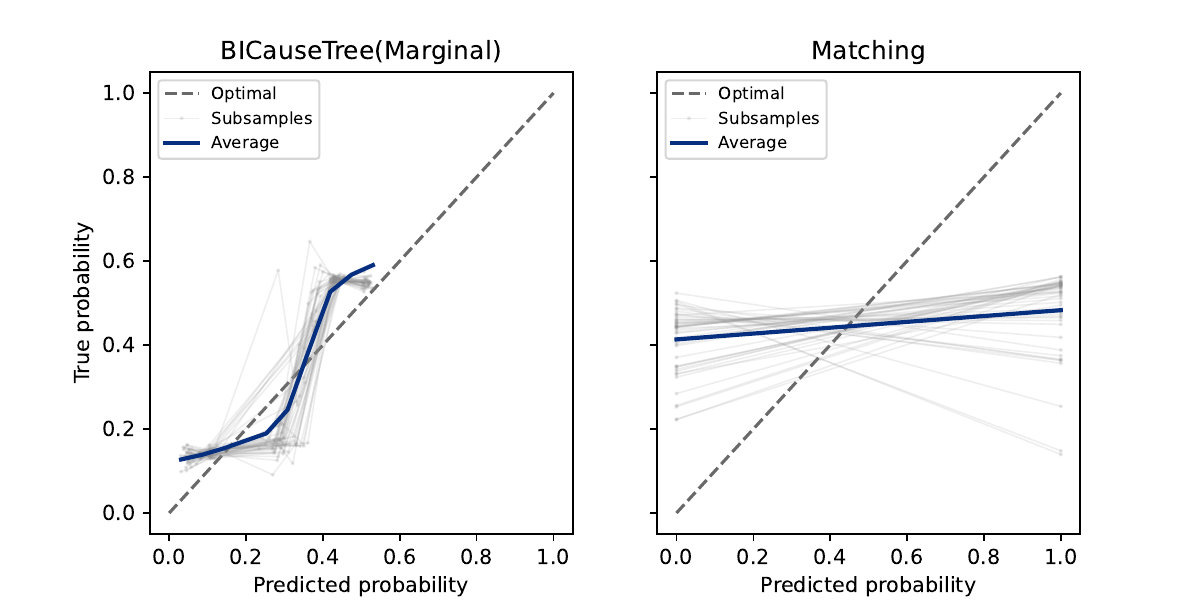}
    \caption{Calibration of the outcome estimation comparing our approach to existing alternatives on the positivity violations testing dataset ($N = 10, 000$) across 50 subsamples}
\end{figure}

\paragraph{Effect estimation bias reduction with tree depth}
We see in this case, that having a max depth that is too large leads to an increase in the bias. This might indicate that a hyperparameter tuning procedure would benefit cases where the tree might become too deep and overfit to the training data.

\begin{figure}[H]
    \centering
    \includegraphics[width=0.5\textwidth]{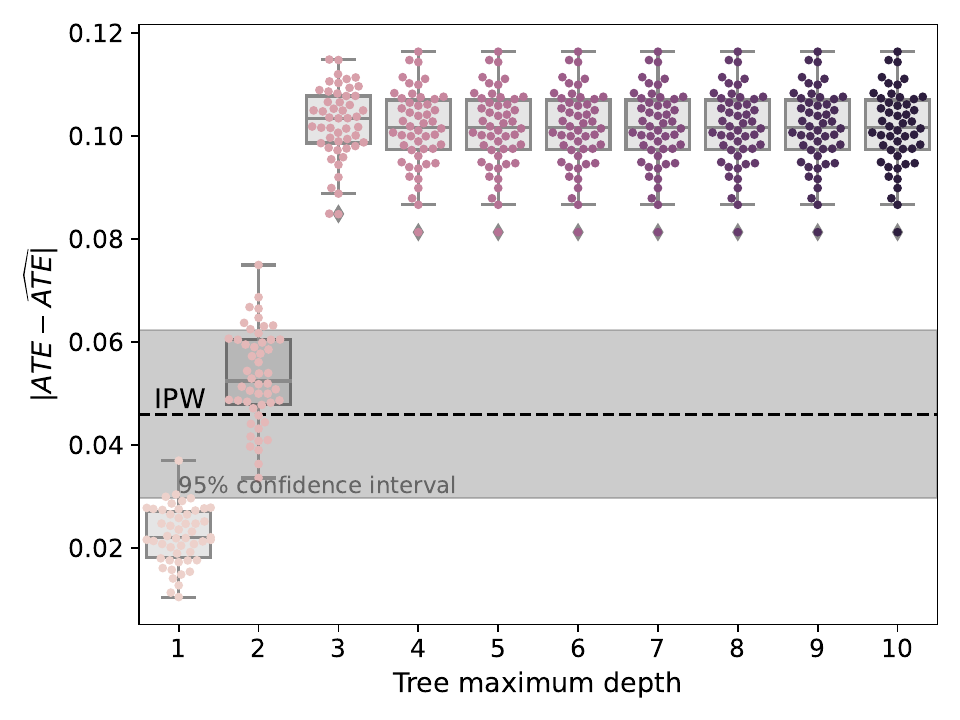}
    \caption{Estimation bias when comparing \textit{BICauseTree(Marginal)} with varying maximum depth parameters with the average bias of IPW (dotted), on the positivity violations experiment training set ($N=10,000$) across 50 subsamples.}
\end{figure}

\paragraph{Treatment allocation bias reduction with tree depth}
Figure \ref{fig:asmd_pos} below shows the weighted ASMD of all three covariates in BICauseTree models with varying maximum tree depth hyperparameters. We see that the ASMD generally decreases with tree depth, showing that the splitting criteria we use leads to balanced subpopulations, as expected. 
\begin{figure}[H]
    \centering
    \includegraphics[width=0.5\textwidth]{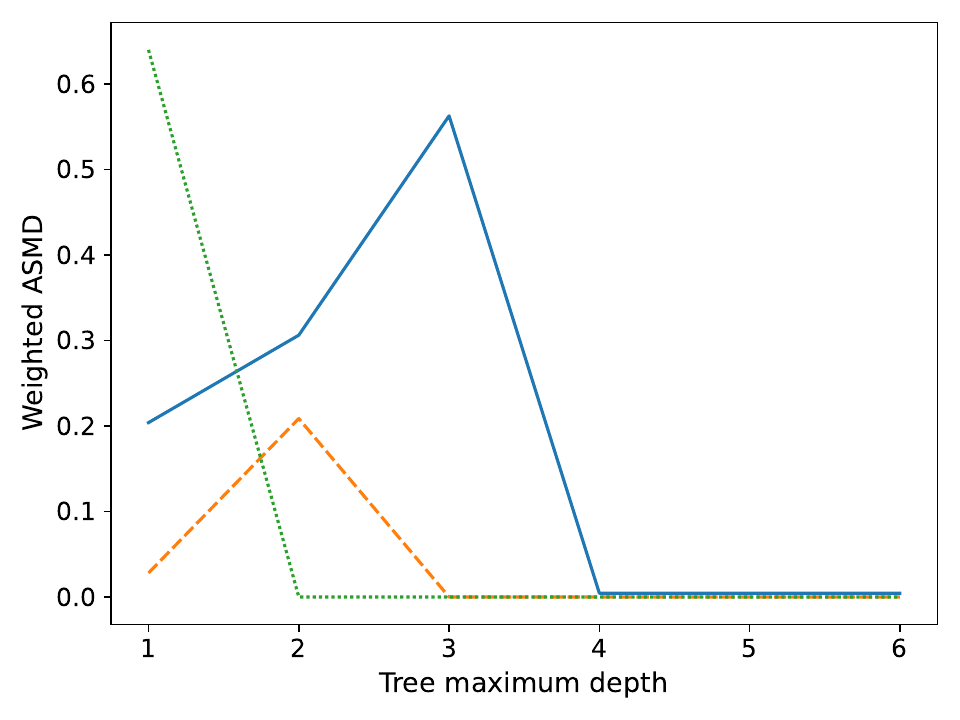}
    \caption{Weighted ASMD for all covariates applying \textit{BICauseTree(Marginal)} models with varying maximum tree depths on the positivity violations dataset training set ($N = 10,000$) across 50 subsamples}
    \label{fig:asmd_pos}
\end{figure}


\clearpage
\newpage

\subsection{Further experiment results: the twins dataset}\label{sec:further_twins}

\paragraph{Partitioning}
The following plot shows the final tree built for the twins dataset. Red nodes indicate positivity violation population.
\begin{figure}[H]
    \centering
    \includegraphics[width=\textwidth]{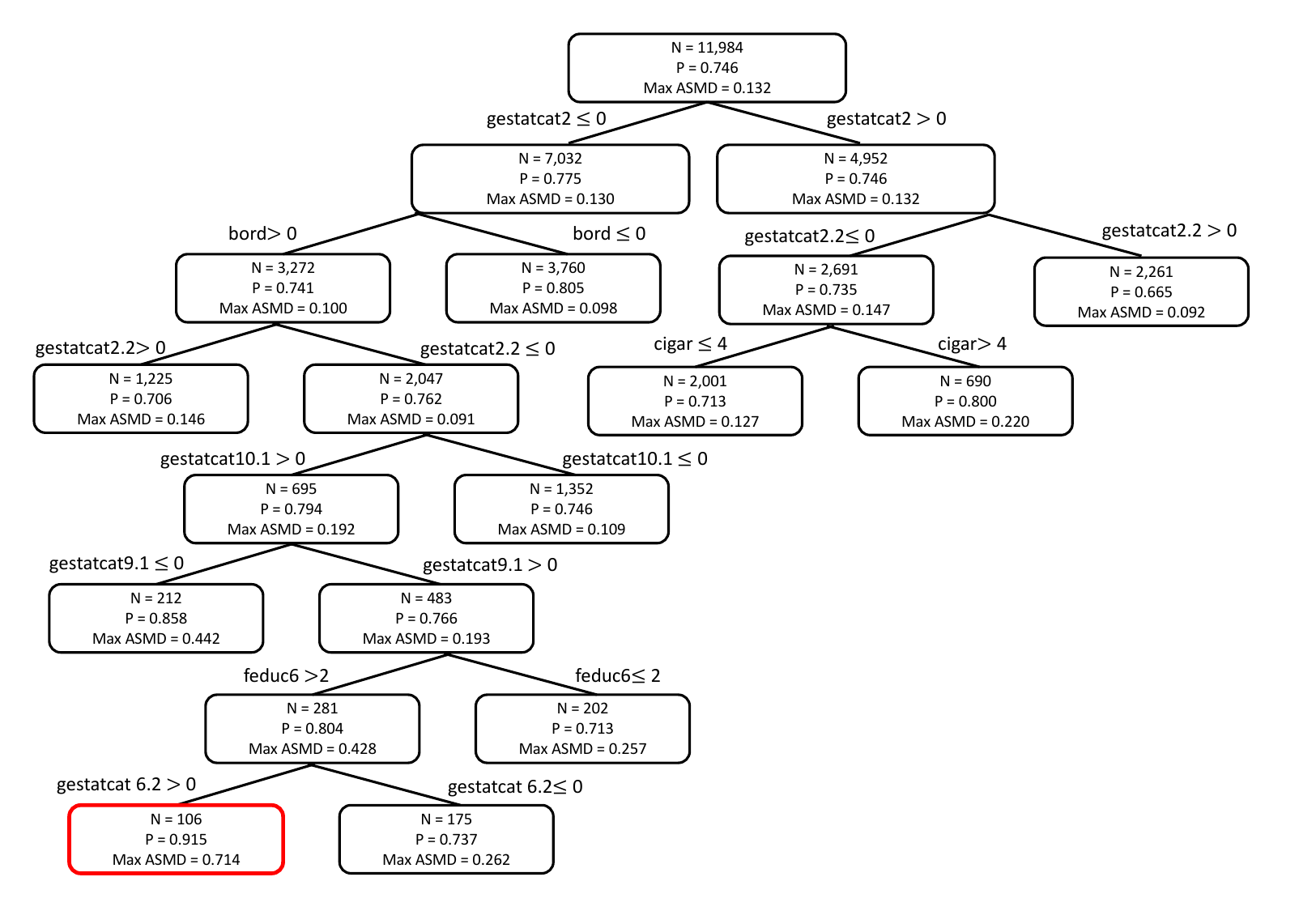}
    \caption{Tree structure after training on the entire twins dataset ($N = 11,984$). Violating leaf nodes are marked in red. P is for node prevalence.}
    \label{fig:tree_entire_twins}
\end{figure}

\paragraph{Causal effect estimation}\label{sec:further_twins_effect}

\begin{figure}[H]
    \centering
    \includegraphics[width=0.5\textwidth]{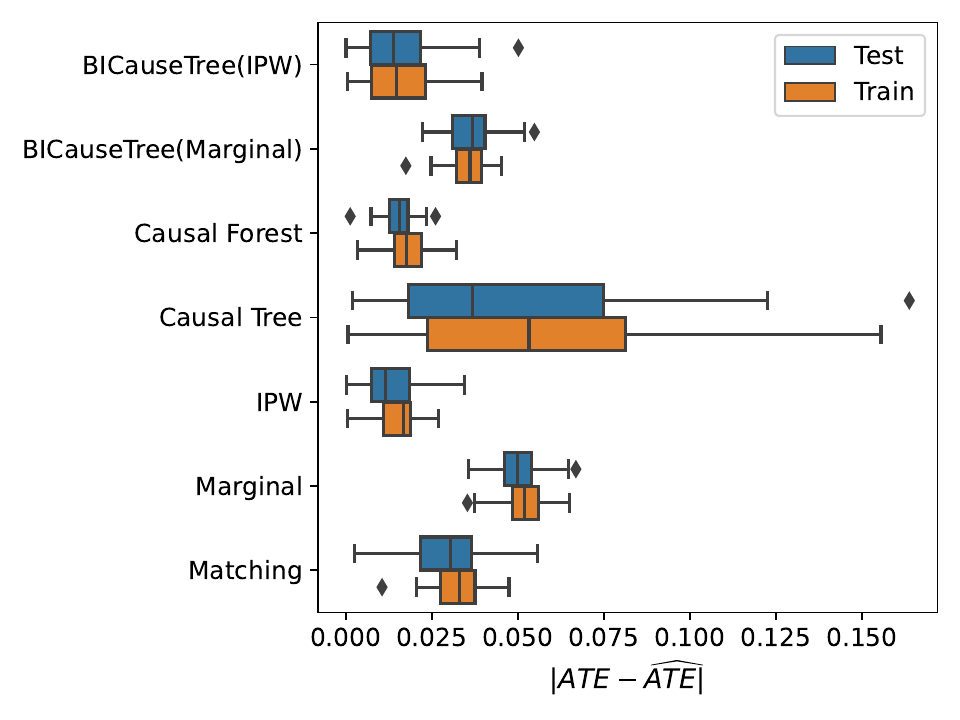}
    \caption{Estimation bias on the twins dataset ($N = 11,984$) across 50 subsamples}
    \label{fig:diff_twins_all}
\end{figure}

Figure \ref{fig:diff_twins_all} shows a comprehensive comparison of the different models. Causal Tree shows poor performance, whereas Causal Forest performs comparably to BICauseTree(Marginal) and IPW.

\paragraph{Outcome estimation}
BICauseTree(Marginal) shows good calibration on outcome estimation. BICauseTree(IPW) seems to have some bias in outcome estimation. This is possibly due to the parametric nature of our IPW model which uses a Logistic Regression. Matching shows poor outcome estimation ability.
\begin{figure}[H]
    \centering
    \includegraphics[width=\textwidth]{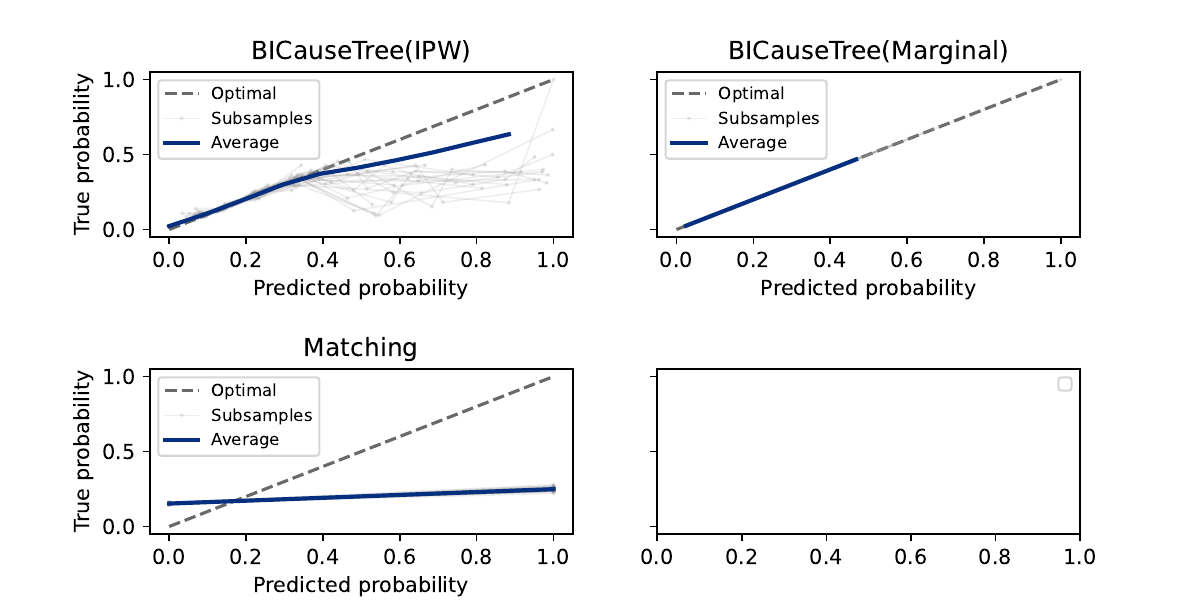}
    \caption{Calibration of the outcome estimation across 50 subsamples, on the twins testing set ($N=5,992$).}
\end{figure}

\paragraph{Treatment allocation bias reduction with tree depth}
Figure \ref{fig:asmd_twins} below shows the reduction in ASMD for the top 10 most imbalanced features in the entire population. It compares BICauseTree models with varying maximum tree depth hyperparameters. In all 10 covariates, we notice how the ASMD reduces as the maximum tree depth increases. 

\begin{figure}[H]
    \centering
    \includegraphics[width=0.5\textwidth]{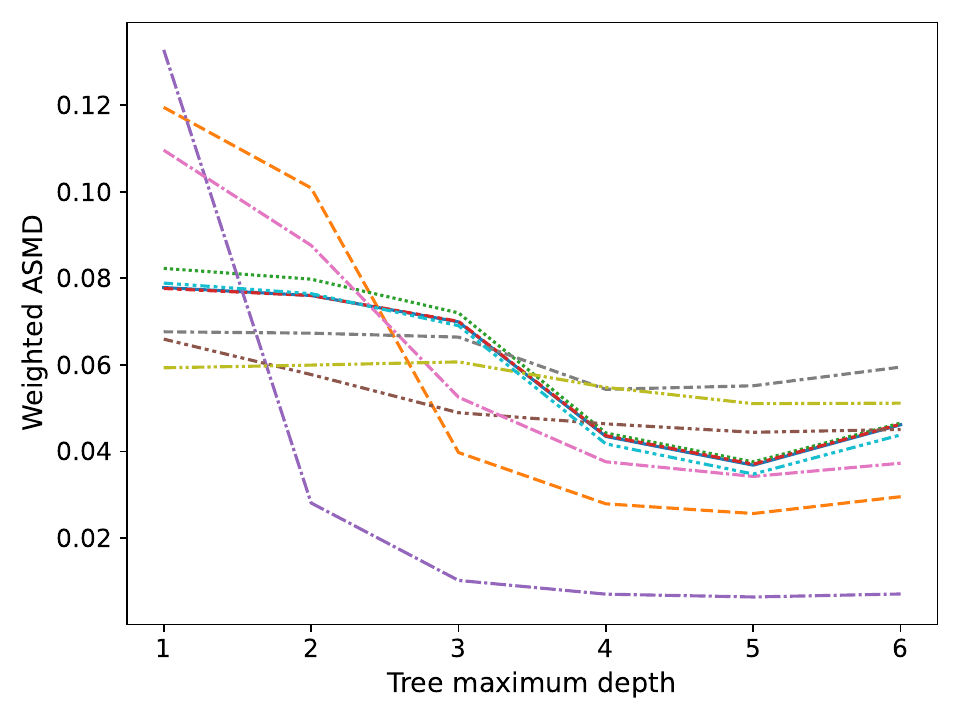}
    \caption{Weighted ASMD for the top 10 covariates with the highest ASMD in the initial population for \textit{BICauseTree(Marginal)} models with varying maximum tree depths on the twins dataset training set ($N = 5,992$) across 50 subsamples}
    \label{fig:asmd_twins}
\end{figure}

\clearpage
\newpage

\subsection{Further experiment results: the ACIC dataset}\label{sec:further_acic}

\paragraph{Partitioning}
The following shows the tree built for the ACIC dataset.
\begin{figure}[H]
    \centering
    \includegraphics[width=.75\textwidth]{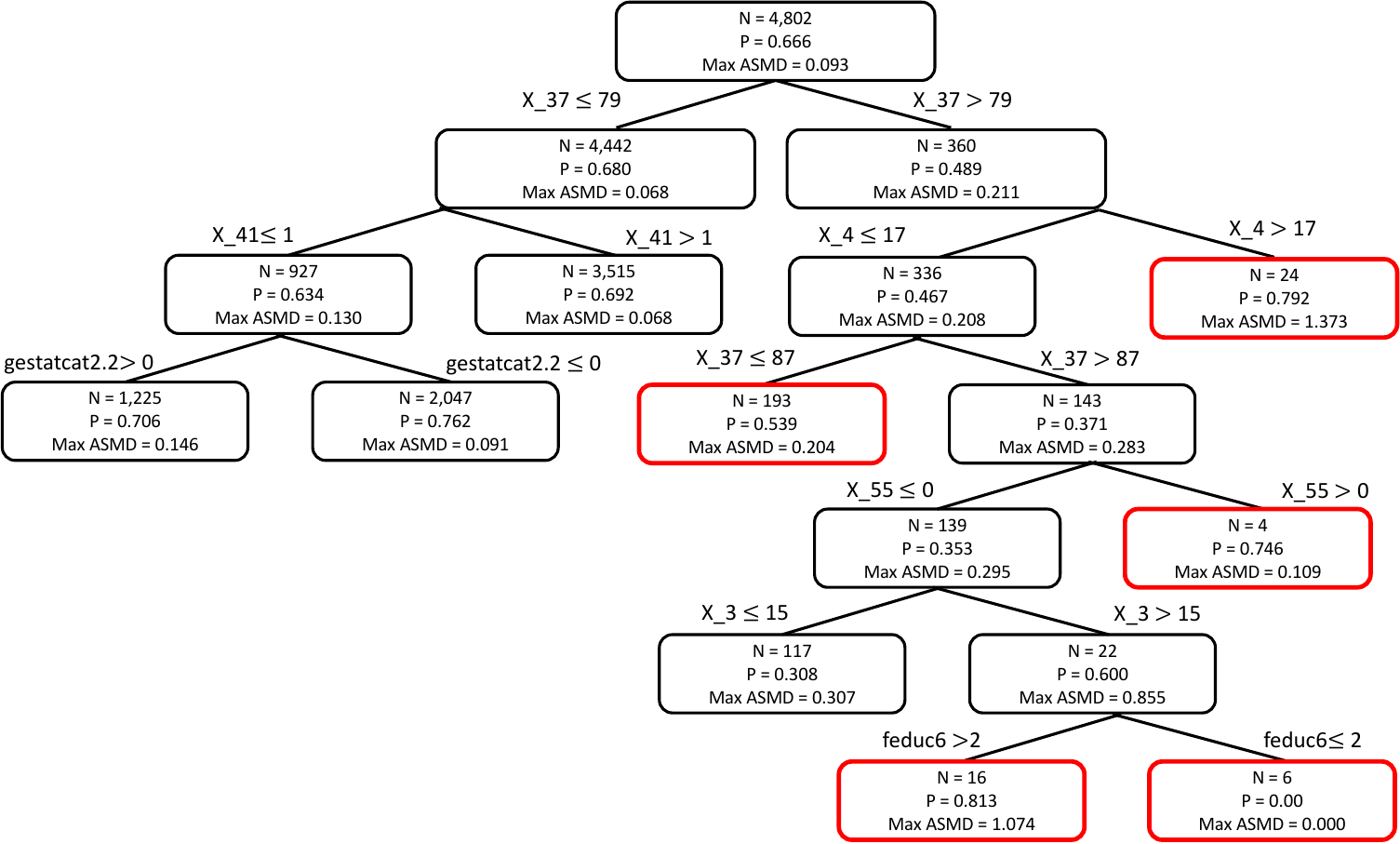}
    \caption{Tree structure after training on the entire ACIC dataset ($N = 4,802$). Violating leaf nodes are marked in red. P is for node prevalence.}
    \label{fig:tree_entire_acic}
\end{figure}

\paragraph{Causal effect estimation}\label{sec:further_acic_effect}
Figure \ref{fig:diff_twins_all} shows a comprehensive comparison of the different models. The performance of both BICauseTree models compares with the one from IPW or Causal Forest. Causal Tree however shows poor performance. Matching shows high estimation bias compared to all other models, including the Marginal estimator. 

\begin{figure}[H]
    \centering
    \includegraphics[width=0.5\textwidth]{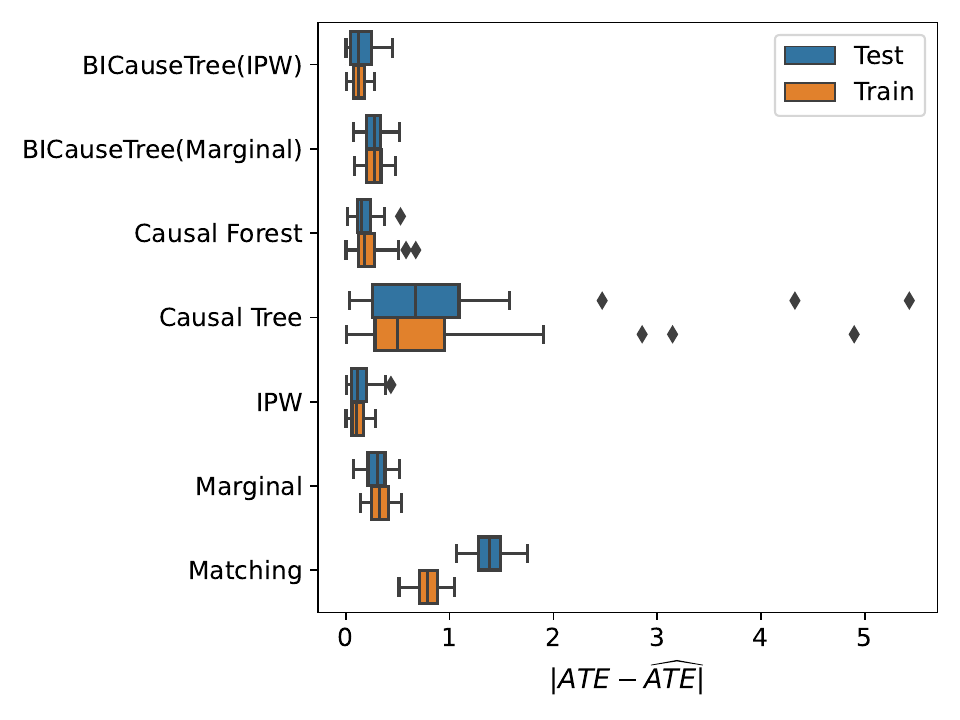}
    \caption{Estimation bias on the ACIC dataset testing set ($N = 960$) across 50 subsamples}
\end{figure}

\paragraph{Propensity score estimation}
We show here the calibration of propensity scores in the ACIC dataset. BICauseTree performs less well than IPW, which is more tailored for propensity estimation. 
As the outcome is non-binary in ACIC, we are not able to generate a calibration plot comparing the outcome estimation of BICauseTree with the one from existing approaches.

\begin{figure}[H]
    \centering
    \includegraphics[width=\textwidth]{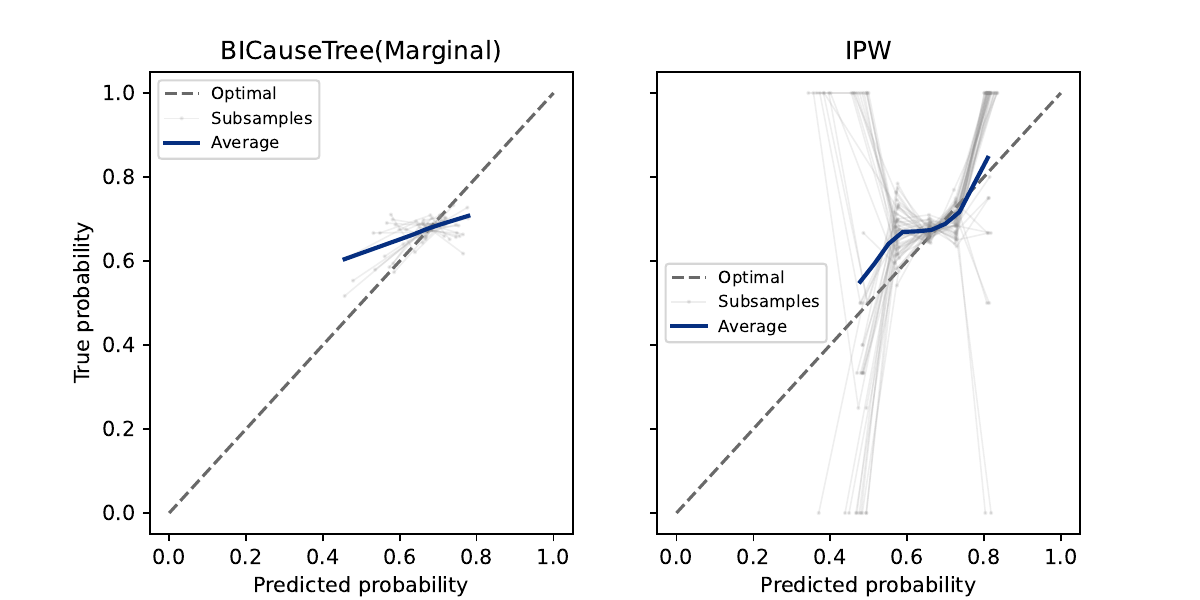}
    \caption{Propensity score calibration comparing our approach to existing alternatives on the ACIC dataset testing set ($N = 960$) across 50 subsamples}
    \label{fig:acic_prop_calibration}
\end{figure}

\paragraph{Effect estimation bias reduction with tree depth}
Figure \ref{fig:bias_depth_acic} below shows the estimation bias for various maximum tree depth hyperparameters. We notice some overlap to the IPW confidence interval, and bias reduces for depths up to max depth 6, then the variance across subsamples starts to increase. This may be due to the limited sample size of this dataset.

\begin{figure}[H]
    \centering
    \includegraphics[width=0.5\textwidth]{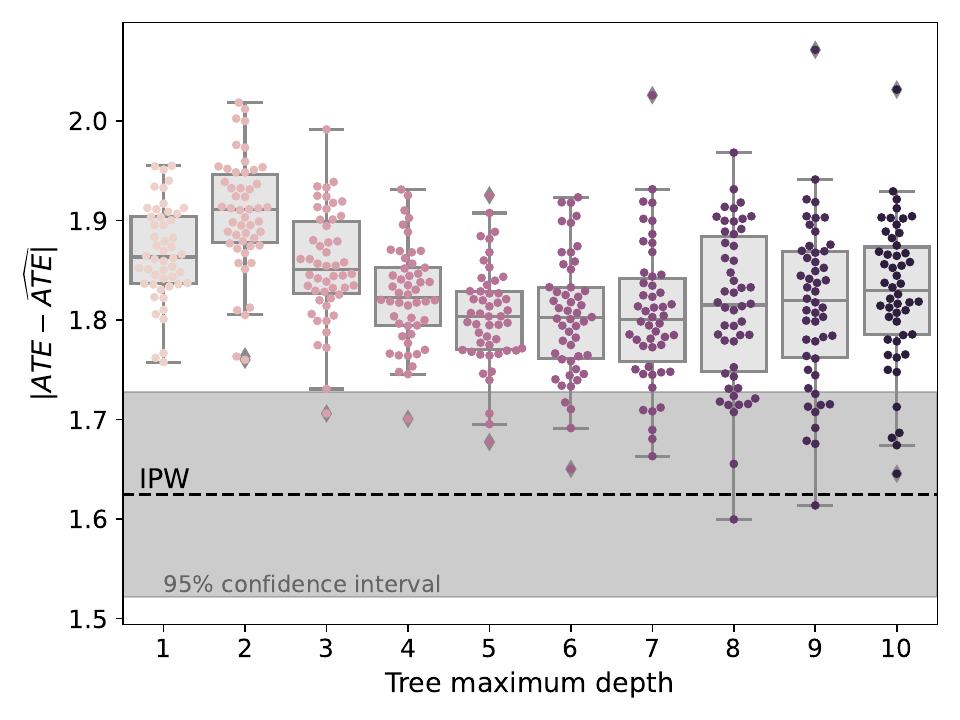}
    \caption{Estimation bias when comparing \textit{BICauseTree(Marginal)} with varying maximum depth parameters with the average bias of IPW (dotted), on the ACIC training set ($N=3,842$) across 50 subsamples.}
    \label{fig:bias_depth_acic}
\end{figure}

\paragraph{Treatment allocation bias reduction with tree depth}
Figure \ref{fig:asmd_acic} below shows the reduction in ASMD for the top 10 most imbalanced features in the entire population. It compares BICauseTree models with varying maximum tree depth hyperparameters. All 10 covariates show a reduction in ASMD with depth.
\begin{figure}[H]
    \centering
    \includegraphics[width=0.5\textwidth]{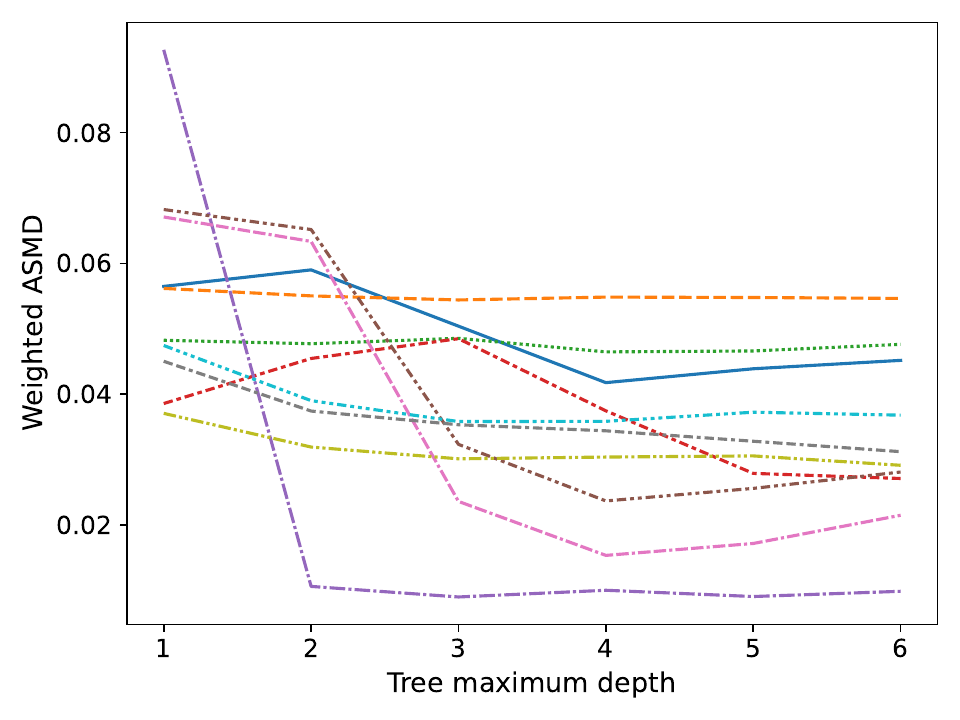}
    \caption{Weighted ASMD across maximum tree depths for the top 10 covariates with the highest ASMD in the initial population, applying \textit{BICauseTree(Marginal)} models with varying maximum tree depths on the ACIC dataset training set ($N = 3,842$) across 50 subsamples}
    \label{fig:asmd_acic}
\end{figure}

\clearpage
\newpage

\subsection{Implementation}\label{sec:implementation_details}

\subsubsection{Positivity violations evaluations}

We implemented two positivity violations definition procedures: the Crump method and a method we introduced, which we call the
\textit{symmetric prevalence} threshold method.
The Crump method is a data-driven method that defines a threshold for extreme propensity scores (i.e., positivity violations), based on the distribution of propensity scores in the data (see further results in \cite{crump2009dealing}). 

The symmetric prevalence procedure generates upper and lower cutoff values that are adjusted for the prevalence. The cutoffs are computed such that if the overall prevalence was 0.5 they would be symmetrical (e.g. for $\alpha=0.05$ the cutoffs would be 0.05 and 0.95). We may consider this as class reweighting of the propensities, within the entire population. 

If we denote the overall prevalence $\mu$, and consider $\alpha$ as the cutoff had the distribution been symmetric (we recommend taking $\alpha=0.1$), the cutoffs are computed as follows:
$$\textit{Upper cutoff} = \frac{(1 - \alpha)*\mu}{(1-\alpha)*\mu + \alpha*(1 - \mu)}$$
$$\textit{Lower cutoff} = \frac{\alpha*\mu}{\alpha*\mu + (1 - \alpha)*(1 - \mu)}$$

\subsection{Experimental details and computation}\label{sec:exp_details}

For BICauseTree, most hyperparameters were set to their default value. Multiple hypothesis test correction was done following a step-down method using Holm-Bonferroni adjustments \cite{holm1979simple, abdi2010holm}, with $\alpha=0.05$. The threshold for weight trimming for positivity violations was computed using the Crump procedure \cite{crump2006moving, crump2009dealing} with 10000 segments. Throughout all experiments the minimum treatment group size was set to 2 patients. The maximum depth is the only parameter which varied depending on the experiments. It was set to 5 for both synthetic datasets and 10 for the experiments on the twins dataset. A long-standing practice has been to define any covariate with $ASMD \geq 0.10$ as a potentially problematic confounder \cite{austin2009balance}, so we set this as our default threshold and used it in our experiments. \\
For IPW(LR), we used a Logistic Regression with a \textit{saga} solver, no penalty and a maximum number of iterations equal to 500. For comparison purposes, BICauseTree(IPW) used similar hyperparameters for its internal IPW outcome model. For IPW(GBT) we used default hyperparameters.\\ 
We applied double matching based on the Mahalanobis distance for all datasets but ACIC, on which we used an Euclidean distance to avoid non-invertable matrices caused by the sparsity of non-null column values. \\
Causal Tree with a single estimator and a subforest size of 1. For Causal Forest, we used 50 estimators and a subforest size of 1.

For including Causal Tree into our experiments, we used the code available at \url{https://github.com/py-why/EconML}. Outcome and propensity models were trained using \verb|sklearn| with default parameters and 500 maximum iterations for Logistic Regression when relevant. Statistical testing was implemented using the \verb|statsmodel| package. 

\subsubsection{Compute and Runtime}

In general, BICauseTree inherits its computational efficiency from decision trees.
Namely the fitting complexity is 
$O\left(M \cdot V \cdot N \log(N) \right)$, 
Where $N$ it the number of observations, 
$M$ is the number of features,
and $V = \max_{m\in M}(V_m)$ the maximal number of unique levels among all $M$ features.
Briefly, recursively splitting the tree is $O(N \log(N))$,
Additionally, for every recursive iteration we consider all $M$ features (calculating ASMD is $O(N)$ for each feature), and, among the select feature we consider a split cutoff over all its unique values (note this is 1 for binary features, calculating a statistical test takes $O(N)$ for each level).
Therefore the resulting complexity is $O\left(M V N \log(N) \right)$.

Table \ref{tab:compute} shows the average runtime it took for each model to fit.
We see BICauseTree compares with IPW, Causal Tree and Causal Forest in terms of wall-clock run-time, while Matching has higher compute time than all other models. 

\begin{table}[h]
    \centering
    \resizebox{0.5\textwidth}{!}{%
    \begin{tabular}{|l|c|}
        \hline
         \textbf{Experiment} & \textbf{Amount of Compute} \\ \hline
         \textbf{Natural experiment dataset} & \\
         BICauseTree(Marginal) & 286  \\
         IPW &  134 \\
         Matching &  882 \\
         Causal Tree & 183 \\
         Causal Forest & 390 \\ 
         \textbf{Positivity violations dataset} &  \\ 
         BICauseTree(Marginal) & 258 \\
         IPW & 128 \\
         Matching & 929 \\ 
         Causal Tree & 137 \\
         Causal Forest & 384 \\ 
         \textbf{Twins} &  \\ 
         BICauseTree(Marginal) & 420 \\
         BICauseTree(IPW) & 567 \\
         IPW & 329 \\
         Matching & 2283 \\
         Causal Tree & 403 \\
         Causal Forest & 672 \\ 
         \textbf{ACIC}
          &  \\
         BICauseTree(Marginal) &  329 \\
         BICauseTree(IPW) & 376 \\
         IPW &  239 \\
         Matching & 1092 \\
         Causal Tree & 354 \\
         Causal Forest & 439 \\ 
         \hline
    \end{tabular}}
    \caption{Total amount of compute in seconds for model fitting across 50 train-test splits, for selected experiments. All experiments were run on a 10 core CPU Apple M1 Pro.}
    \label{tab:compute}
\end{table}

\begin{figure}[H]
    \centering
    \includegraphics[width=0.5\linewidth]{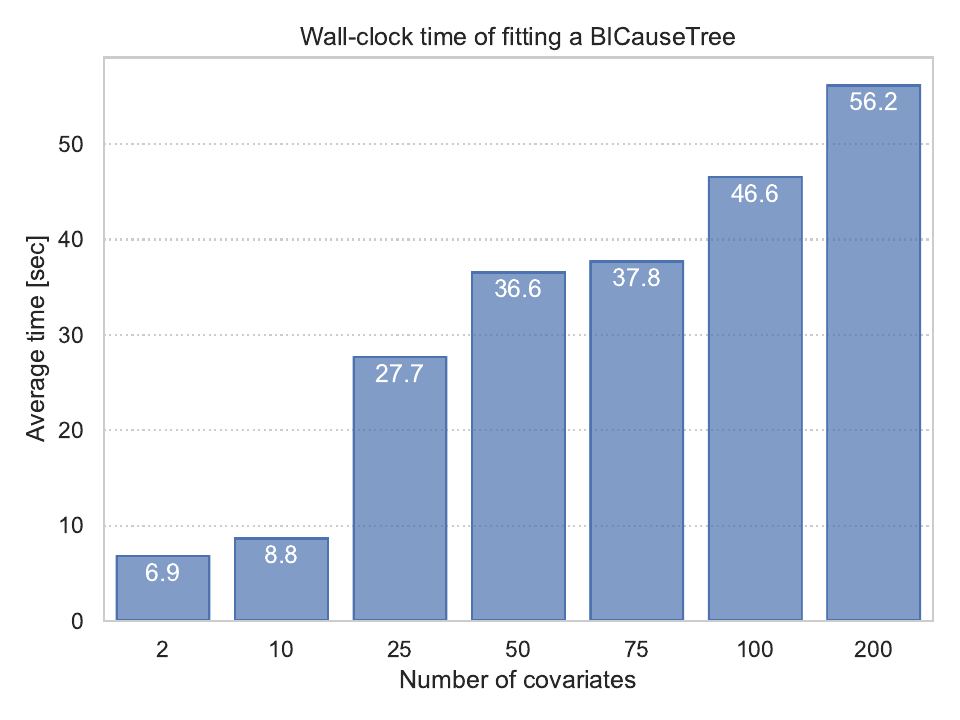}
    \caption{Runtime of BICauseTree as a function of increasing the dimensionality of the input. The bars show the average wall-clock time, in seconds, it took to fit a BICauseTree for different numbers of input covariates.}
    \label{fig:high_dim_time}
\end{figure}

We further tested the runtime of BICauseTree as a function of dimensionality.
Figure \ref{fig:high_dim_time} shows the time, averaged over 10 random splits, it took for BICauseTree to fit---including filtering for positivity---as a function of the number of covariates in the input data (following the experiment design presented earlier in the Appendix in Figure \ref{fig:high_dim_bias}.

\subsubsection{Causal benchmark datasets}
The \textbf{twins dataset} was originally taken from the denominator file at \url{https://www.nber.org/research/data/linked-birthinfant-death-cohort-data}. However, we use data generated by \textit{Neal et. al} \cite{neal2020realcause}, which simulates an observational study from the initial data by selectively hiding one of the twins with a generative approach. The sample size is $N = 11,984$ pairs of twins, with the essential inclusion criterion being that both individuals were born weighing less than 2kg. The mortality rate amongst the lighter twins is 18.9\%, and for the heavier 16.4\%, for an average treatment effect of $-2.5\%$ (which we thus consider as ground truth). A total of 75 covariates were recorded, relating to the parents' socio-demographic features and medical history, the pregnancy and the birth.

The \textbf{ACIC dataset} was generated using the \textit{causallib} package at \url{https://github.com/BiomedSciAI/causallib/blob/master/causallib/datasets/data/acic_challenge_2016/README.md}. It contains covariates, simulated treatment, and simulated response variables for the causal inference challenge in the 2016 Atlantic Causal Inference Conference \cite{dorie2019automated}. For each of 20 conditions, treatment and response data were simulated from real-world data corresponding to 4802 individuals and 58 covariates. After one-hot encoding, a total of 79 covariates was included. More specifically, we used the set of treatment and response variables \textit{zymu 174570858}, with the two expected potential outcomes (\textit{mu0}, \textit{mu1}).

\subsection{Social impact of our work: further details}\label{sec:social_impact}

Recent years have seen a surge in the Explainable AI (XAI) literature, motivated by rising ethical concerns around artificial intelligence. Model scrutiny is particularly relevant in sensitive environments such as healthcare, which require high safety standards considering the major consequences of invalid predictions on individual trajectories \cite{yampolskiy2018artificial}. Calls for model transparency are further motivated by evidences that supervised machine learning is inclined to reproduce inherent bias and prejudice against discriminated groups \cite{challen2019artificial}. Ultimately, XAI aims at building trust in models that ought to be deployed. In recent years, the demand for transparency expanded beyond the research community, notably with incentives from high institutions. The European Union General Data Protection Regulation legislation has mandated a ``right to explanation'' for individual predictions that can ``significantly affect'' users \cite{holzinger2017we}. In other words, algorithmic results should be re-traceable on demand. In parallel, quality control frameworks such as the U.S. Food and Drug Administration guidance have been introduced to ensure the safety of clinical AI \cite{benjamens2020state}. 
By providing interpretable effect estimation, our BICauseTree approach aligns with the mission of increasing transparency for downstream users. As such, it is most likely to have positive social impact.  We however caution against relying exclusively on causal inference when data accuracy or volumes are insufficient, as misleading effect estimates may have negative social impact.

\end{document}